\documentclass[aps,prd,onecolumn,eqsecnum,amsmath,nofootinbib,preprintnumbers]{revtex4}%

\usepackage{color,graphicx,float,subfigure,xcolor}%[dvips]%
\usepackage{amsfonts,amssymb,theorem,mathrsfs,times}
\usepackage{bm}
\usepackage{mathtools}
\usepackage{amsfonts,amssymb,theorem,mathrsfs}
\usepackage{dsfont}
\usepackage{setspace}
\usepackage{amsmath}  
\usepackage[colorlinks,linkcolor=blue,anchorcolor=blue,citecolor=blue]{hyperref}   % a package related to hyperlink
\usepackage{ulem}   % a package related to strikethrough
\usepackage[english]{babel}
\usepackage{tikz}

\usepackage{accents}
\usepackage{url}

{\theorembodyfont{\upshape}
	}
{\theorembodyfont{\upshape}
	}
{\theorembodyfont{\upshape}
	}
{\theorembodyfont{\upshape}
	}
{\theorembodyfont{\upshape}
	}
{\theorembodyfont{\upshape}
	}
\addtocounter{MaxMatrixCols}{10}
\newcommand{\dalm}{\kern1pt\vbox{\hrule height 0.9pt\hbox{\vrule width
			0.9pt\hskip 2.5pt\vbox{\vskip 5.5pt}\hskip 3pt\vrule width
			0.3pt}\hrule height 0.3pt}\kern1pt}

\begin{document}
\thispagestyle{empty}
\preprint{\hfill {\small {ICTS-USTC/PCFT-25-02}}}
%<<<<<<<<<<<<< TITLE >>>>>>>>>>>>>>>%
\title{The pseudospectrum for the Kerr black hole with spin $s=0$ case}
	
%<<<<<<<<<<<<< AUTHOR >>>>>>>>>>>>>>>%
%\author{$^b$}
%
%\email{}
\author{ Rong-Gen Cai$^{d\, ,a\, ,b}$\footnote{e-mail address: cairg@itp.ac.cn}} 

\author{ Li-Ming Cao$^{e\, ,f}$\footnote{e-mail address: caolm@ustc.edu.cn}} 

\author{ Jia-Ning Chen$^{a\, ,b\, ,c}$\footnote{e-mail address: chenjianing22@mails.ucas.ac.cn}}

\author{ Zong-Kuan Guo$^{b\, ,c\, ,a}$\footnote{e-mail address: guozk@itp.ac.cn}}

\author{ Liang-Bi Wu$^{a\, ,c}$\footnote{e-mail address: liangbi@mail.ustc.edu.cn (corresponding author)} }

% \author{ Libo Xie$^a$\footnote{e-mail address: xielibo23@mails.ucas.ac.cn}}

% \author{ Long-Yue Li$^c$\footnote{e-mail
% 			address: lily26@mail.ustc.edu.cn}}

 \author{Yu-Sen Zhou$^f$\footnote{e-mail address: zhou\_ys@mail.ustc.edu.cn}}	
	
%<<<<<<<<<<<<< ADDRESS >>>>>>>>>>>>>>>%
\affiliation{${}^a$School of Fundamental Physics and Mathematical Sciences, Hangzhou Institute for Advanced Study, UCAS, Hangzhou 310024, China}

\affiliation{${}^b$CAS Key Laboratory of Theoretical Physics, Institute of Theoretical Physics, Chinese Academy of Sciences, Beijing 100190, China}

\affiliation{${}^c$University of Chinese Academy of Sciences, Beijing 100049, China}

\affiliation{${}^d$Institute of Fundamental Physics and Quantum Technology, Ningbo University, Ningbo 315211, China}

\affiliation{${}^e$Peng Huanwu Center for Fundamental Theory, Hefei, Anhui 230026, China}

\affiliation{${}^f$Interdisciplinary Center for Theoretical Study and Department of Modern Physics, University of Science and Technology of China, Hefei, Anhui 230026, China}

%<<<<<<<<<<<<< DATE >>>>>>>>>>>>>>>%
\date{\today}
	
%======================================%
%<<<<<<<<<<<<< ABSTRACT >>>>>>>>>>>>>>>%
%======================================%
\begin{abstract}
We investigate the pseudospectrum of the Kerr black hole, which indicates the instability of the spectrum of quasinormal modes (QNMs) of the Kerr black hole. Methodologically, we use the hyperboloidal framework to cast the QNM problem into a two-dimensional eigenvalue problem associated with a non-self-adjoint operator, and then the spectrum and pseudospectrum are solved by imposing the two-dimensional Chebyshev collocation method. The (energy) norm is constructed by using the conserved current method for the spin $s=0$ case. For the finite rank approximation of the operator, we discuss the convergence of pseudospectra using various norms, each involving different orders of derivatives. The convergence of the pseudospectrum improves as the order of the derivatives increases. We find that an increase in the imaginary part of complex frequency can deteriorate the convergence of the pseudospectrum under the condition of the same norms.

\end{abstract}

\maketitle
	
%======================================%
%<<<<<<<<<<< Introduction >>>>>>>>>>>>>%
%======================================%	
\section{Introduction}
Based on general relativity (GR), the collision of two black holes leads to the creation of a highly distorted black hole, which subsequently transitions into a Kerr black hole via a ringdown phase. The gravitational waves (GWs) produced during the ringdown period are considered to be accurately modeled by the linear perturbation theory of black holes, under the guidance of the Teukolsky equation (TE)~\cite{Teukolsky:1972my,Teukolsky:1973ha}. With the radiation boundary condition or the Sommerfeld condition, TE has the quasinormal mode (QNM) solutions. An important property of the spectra of QNMs is that they are only determined by the mass and spin of the Kerr black hole. Using these QNM spectra and digital signal processing (DSP) techniques of gravitational waves, one can estimate the mass and spin of black holes based on the detected gravitational wave data~\cite{Baibhav:2023clw,Giesler:2019uxc}.

Due to the fact that BHs are not isolated and they are always surrounded by some matters, the QNM spectrum of such BHs affected by the astrophysical environment has gained considerable attention as the improvement of the accuracy of the GWs detection~\cite{Barausse:2014tra}. The QNM spectra of black holes exhibit an instability, causing them to shift disproportionately far in the complex plane in response to seemingly minor environmental perturbations. Initial studies of QNM spectrum instability was provided by Nollert and Price~\cite{Nollert:1996rf,Nollert:1998ys}. There are two main categories of methods used to study spectrum instability in the aspect of QNMs. The first approach involves modifying the effective potential based on various reasons, which can be either physically motivated or artificially constructed~\cite{Daghigh:2020jyk,Qian:2020cnz,Berti:2022xfj,Cheung:2021bol,Li:2024npg,Yang:2024vor,Courty:2023rxk,Cardoso:2024mrw,Ianniccari:2024ysv}.  Consider that the instability of the QNM spectrum, several stable physical quantities have been studied, such as greybody factor~\cite{Oshita:2023cjz,Rosato:2024arw,Oshita:2024fzf,Wu:2024ldo}, scattering cross section and absorption cross section~\cite{Torres:2023nqg}.

Pseudospectrum analysis~\cite{trefethen2020spectra}, as a second approach, is utilized to investigate the instability of the QNM spectrum~\cite{Boyanov:2024fgc, Jaramillo:2020tuu,Destounis:2023ruj,Jaramillo:2021tmt}, which originally comes from the field of hydrodynamics~\cite{science.261.5121.578}. This method involves analyzing the properties of non-self-adjoint operators in dissipative systems and provides a visual understanding of the instability in the spectrum of such operators. In the context of black holes, pseudospectra have been employed to identify qualitative characteristics that serve as indicators of spectrum instability across a range of spacetimes, including asymptotically flat black holes~\cite{Jaramillo:2020tuu,Destounis:2021lum,Cao:2024oud,Cao:2024sot}, asymptotically AdS black holes~\cite{Arean:2024afl,Garcia-Farina:2024pdd,Arean:2023ejh,Boyanov:2023qqf,Cownden:2023dam}, asymptotically dS black holes~\cite{Sarkar:2023rhp,Destounis:2023nmb,Luo:2024dxl,Warnick:2024usx}, and horizonless compact objects~\cite{Boyanov:2022ark}. Transient dynamics related to pseudospectra are studied in~\cite{Carballo:2024kbk,Jaramillo:2022kuv,Chen:2024mon}. More recently, a hyperboloidal Keldysh's approach in terms of black hole QNM problems is put forward by~\cite{Besson:2024adi}. 

The pseudospectrum ought to be actively integrated into the black hole spectroscopy program, and a thorough examination of the pseudospectrum of black holes is essential for advancing our understanding of spectrum stability. The basic method for studying the pseudospectrum is the hyperboloidal approach of QNMs~\cite{PanossoMacedo:2024nkw}. In traditional representation, QNM eigenfunctions exhibit divergent behavior toward the bifurcation sphere $\mathcal{B}$ and spatial infinity $i^0$. In contrast, when assessed in proximity to the future event horizon $\mathcal{H}^+$ and at null infinity $\mathscr{I}^+$, QNM eigenfunctions exhibit regularity. Hyperboloidal surfaces serve as a natural bridge between these domains, offering a geometric framework to regularize QNMs. More about the hyperboloidal approach can be found in~\cite{Zenginoglu:2007jw,Zenginoglu:2011jz,PanossoMacedo:2023qzp,Zenginoglu:2024bzs}.

However, while the previous works are focusing on the case of spherical symmetry, the case of non-spherical symmetry remains an open question. In this paper, we study the pseudospectrum for the Kerr black hole. This endeavor is vital for determining whether QNM spectrum instabilities are a feature of spherical symmetry alone or an ubiquitous trait of all compact objects in the universe. The hyperboloidal framework for the Kerr black hole was given by~\cite{PanossoMacedo:2019npm}, which is regarded as our basic guideline. Recently, for the Kerr black hole, Ref.\cite{Minucci:2024qrn} provides a geometrical interpretation of the confluent Heun functions (Note that Ref.\cite{Chen:2023ese} uses the confluent Heun functions to get the exact solutions of TE.) within black hole perturbation theory and elaborates on their relation to the hyperboloidal framework. From such a hyperboloidal framework, TE can be transformed into a hyperbolic partial differential equation written as $\partial u/\partial\tau=iLu$, where $\tau$ refers to the time coordinate and the operator $L$ is called an infinitesimal time generator. The QNM boundary conditions have been incorporated into the operator $L$. Using the two-dimensional pseudo-spectral method (spectral collocation method), one can further transform the obtained system of partial differential equations into a system of ordinary differential equations, which can be symbolically written as $\mathrm{d}\mathbf{u}/\mathrm{d}\tau=i\mathbf{L}\mathbf{u}$ with $\mathbf{L}$ being time-independent. After performing a Fourier transformation, the QNM frequency $\omega$ can be solved as an eigenvalue problem for the matrix equation $\mathbf{L}\mathbf{u}=\omega\mathbf{u}$, which is a two-dimensional eigenvalue problem. 

Recently, there have been many studies on two-dimensional QNM spectrum problems~\cite{Blazquez-Salcedo:2023hwg,Chung:2023wkd,Khoo:2024yeh,Chung:2024ira,Chung:2024vaf,Blazquez-Salcedo:2024oek,Blazquez-Salcedo:2024dur,Ripley:2022ypi} avoiding variable separations. All of these works are based on the spectral decomposition of metric perturbations. A significant advantage lies in the elimination of the need for variable separation in the derived perturbation equations, which facilitates the application of this method to various modified theories of gravity, such as Einstein-scalar-Gauss-Bonnet (EsGB) gravity theory. An advantage of the spectral collocation method over the spectral decomposition method lies in the fact that we can not only obtain the QNM spectrum, but also transform the operator describing QNM dynamics into a matrix operator (finite rank approximation). The pseudospectrum of the original operator can be approximated by the pseudospectrum of this matrix operator~\cite{trefethen2020spectra}. 

The establishment of pseudospectrum is based on how to define the norm, thus the selection of a norm becomes a pertinent issue when discussing the pseudospectrum. Different norms will lead to different pseudospectra, therefore there is an interest question that whether the different pseudospectra would like to give different insights in QNM spectrum instability or not. Specifically, the pseudospectrum can be characterized in terms of the norm of the resolvent $R_L(\omega)=(L-\omega \mathbb{I})^{-1}$, i.e., $\lVert R_L(\omega)\rVert$, where $\omega\in\mathbb{C}$ is referred to frequency. Based on this motivation, three types of norms, which are energy norm, $L^{2}$-norm, and $H^{k}$-norm~\cite{Boyanov:2023qqf,Besson:2024adi}, are used to investigate the impact of norm on QNM instability in our study. The first two types norms are a general choice for many studies, the $H^{k}$-norm is first introduced in the Schwarzschild-AdS black hole~\cite{Boyanov:2023qqf}, mainly considering the regularity of QNMs~\cite{Warnick:2013hba}. Refs.\cite{Jaramillo:2020tuu,Gasperin:2021kfv} tell us that the physically motivated norm associated with the QNM problem is the energy norm coming from the conserved current method at least for the spherically symmetric situation. In this paper, we apply the conserved current method with a timelike Killing vector to the case of Kerr black holes and obtain a relevant energy norm. It is expected that such an energy norm is an $H^1$-norm. Furthermore, we obtain the $H^{k}$-norm by adding a series of higher-order derivative terms in the energy norm. So far, regarding gravity theory, it seems that people can only use numerical methods to calculate the pseudospectrum of the original operator by computing the pseudospectrum of the corresponding matrix operator. Given this, what we need to focus on is the convergence of the pseudospectrum of the matrix operator, i.e., how the norm of its resolvent changes with resolutions.

This paper is organized as follows. In Sec.\ref{TE_QNMs}, we show how TE can be written in $\partial u/\partial\tau=iLu$ and get the QNM spectra. In Sec.\ref{norm}, we construct several norms, containing $L^2$-norm, energy norm, and $H^k$-norm, which will be used in our work. In Sec.\ref{pseudospectrum_and_convergence}, we show the pseudospectrum of Kerr black hole for the spin $s=0$ case and discuss the convergence of the pseudospectrum for the finite rank approximation. Sec.\ref{sec: conclusions} is the conclusions and discussion. In addition, there are four appendices. Appendix \ref{expressions_L1_and_L2} shows the explicit expression of the operator $L$. Appendix \ref{two_dimensional_pseudo_spectral_method} introduces the two-dimensional pseudospectral methods. Appendix \ref{Gram_matrix} is used to construct the Gram matrix using the Clenshaw-Curtis quadrature. In Appendix \ref{other_alpha}, the pseudospectrum for the other dimensionless spin $\alpha$ is given with different norms.

\section{Teukolsky equation and quasinormal modes}\label{TE_QNMs}
We begin this section by reviewing the metric of the Kerr black hole in the Boyer-Lindquist coordinates $(t,r,\theta,\varphi)$
\begin{eqnarray}
    \mathrm{d}s^2=-\Big(1-\frac{2Mr}{\Sigma}\Big)\mathrm{d}t^2-\frac{4Mar}{\Sigma}\sin^2\theta\mathrm{d}t\mathrm{d}\varphi+\frac{\Sigma}{\Delta}\mathrm{d}r^2+\Sigma\mathrm{d}\theta^2+\sin^2\theta\Big(\Sigma_0+\frac{2Ma^2r}{\Sigma}\sin^2\theta\Big)\mathrm{d}\varphi^2\, ,
\end{eqnarray}
where
\begin{eqnarray}
    \Delta(r)&=&r^2-2Mr+a^2=(r-r_{+})(r-r_{-})\, ,\nonumber\\
    \Sigma(r,\theta)&=&r^2+a^2\cos^2\theta\, ,\quad \Sigma_0(r)=\Sigma(r,0)=r^2+a^2\, ,
\end{eqnarray}
and  $M$ and $a$ are the mass and the angular momentum parameters of the Kerr black hole. $r_{+}$ and $r_{-}$ are defined as the event horizon and the Cauchy horizon of Kerr black hole. Then, the Teukolsky equation (TE) in the the Boyer-Lindquist
coordinates $(t,r,\theta,\varphi)$ reads~\cite{Teukolsky:1972my,Teukolsky:1973ha}
\begin{eqnarray}\label{BL_TE}
    0&=&\Big[\frac{(\Sigma_0)^2}{\Delta}-a^2\sin^2\theta\Big]\partial^2_{tt}\Psi^{(s)}+\frac{4Mar}{\Delta}\partial^2_{t\varphi}\Psi^{(s)}+\Big[\frac{a^2}{\Delta}-\frac{1}{\sin^2\theta}\Big]\partial^2_{\varphi\varphi}\Psi^{(s)}\nonumber\\
    &&-\Delta^{-s}\partial_r\Big(\Delta^{s+1}\partial_r\Psi^{(s)}\Big)-2s\Big[\frac{M(r^2-a^2)}{\Delta}-(r+ia\cos\theta)\Big]\partial_t\Psi^{(s)}\nonumber\\
    &&-2s\Big[\frac{a(r-M)}{\Delta}+i\frac{\cos\theta}{\sin^2\theta}\Big]\partial_{\varphi}\Psi^{(s)}-\frac{1}{\sin\theta}\partial_{\theta}\Big(\sin\theta\partial_\theta\Psi^{(s)}\Big)+s(s\cot^2\theta-1)\Psi^{(s)}\, .
\end{eqnarray}
In the above TE, $s$ is the spin-weight parameter. The scalar, electromagnetic, and gravitational perturbations are described by $s=0$, $s=\pm1$ and $s=\pm2$, respectively. A very simple covariant form of the TE is given by~\cite{Bini:2002jx}
\begin{eqnarray}\label{covariant_form_TE}
    \Big[(\nabla^a+s\Gamma^a)(\nabla_a+s\Gamma_a)-4s^2\Psi_2\Big]\Psi^{(s)}=0\, ,
\end{eqnarray}
where $\Gamma^a$ denotes the ``connection vector", and their expressions can be found in~\cite{Bini:2002jx}.

The Boyer-Lindquist coordinate $(t,r,\theta,\varphi)$ is widely used due to the simplicity of the resulting equations, but the disadvantage is the need to impose external boundary conditions to describe a physical scenario composed of a black-hole horizon and a radiation zone. For example, the gravitational self-force (GSF) approach relies on the construction of a retarted potential, the external boundary conditions must be imposed in terms of the retarted $u \sim t-r$ or advanced time $v \sim t+r$. However, at the second order, this approach will lose accuracy at late times as well as large distances~\cite{Pound:2015wva}. Therefore, it is necessary to construct a comprehensive and methodical framework to adapt the time coordinate to the geometric configuration of spatial scales, both in the vicinity of the black hole and at the distant radiation zone~\cite{PanossoMacedo:2019npm}. The hyperboloidal framework comes into being.
        
The hyperboloidal framework removes the necessity of imposing external boundary conditions, since the time coordinate is naturally adapted to the causal structure of the black hole and the radiation zone.  When solving the QNM problem, the QNM boundary conditions are built into the ``bulk'' of the operator, and no additional boundary conditions are required. The light cones point outward at the boundary of the computation domain, simplifying the boundary conditions to only requiring a regular solution, which is trivially satisfied in numerical calculations~\cite{PanossoMacedo:2024nkw}.

For the Kerr black hole, the hyperboloidal framework is completed by Macedo~\cite{PanossoMacedo:2019npm}. The complete mapping from Boyer-Lindquist to the hyperboloidal coordinates are given by~\cite{PanossoMacedo:2019npm}
\begin{eqnarray}\label{coordinate_transformation}
    t=\lambda\Big[\tau-h(\sigma,\theta)\Big]-r_{\star}(r(\sigma))\, ,\quad r(\sigma)=\lambda\frac{\rho(\sigma)}{\sigma}\, ,\quad \varphi=\phi-k(r(\sigma))\, .
\end{eqnarray}
Here, $\lambda$ is a length scale of the black hole and $h(\sigma,\theta)$ is called the height function. The height function $h(\sigma,\theta)$ and the radial function $\rho(\sigma)$ characterize the degrees of freedom of the gauge. The expressions of them depend on different kinds of guages. As for the so-called minimal gauge, its subdivisions are the radial function fixing gauge used, which will be used in this study, and the Cauchy horizon fixing gauge~\cite{PanossoMacedo:2019npm}. For more gauges, one can refer to Ref.\cite{Zenginoglu:2007jw} and Ref.\cite{PanossoMacedo:2019npm}. Furthermore, the tortoise $r_{\star}(r)$ and the phase $k(r)$ are defined by
\begin{eqnarray}
    \frac{\mathrm{d}r_{\star}}{\mathrm{d}r}=\frac{\Sigma_0}{\Delta}\, ,\quad \frac{\mathrm{d}k}{\mathrm{d}r}=\frac{a}{\Delta}\, .
\end{eqnarray}
For the hyperboloidal coordinates $(\tau,\sigma,\theta,\phi)$, the black hole event horizon is at $\sigma=\sigma_{+}$, and the future null infinity is at $\sigma=0$.

In order to regularize the essential singularities in the radial direction at future null infinity $\sigma=0$ and the black hole horizon $\sigma=\sigma_{+}$, and to realize the essential regularity at $\sin\theta=0$, the master function $\Psi^{(s)}$ of Eq.(\ref{BL_TE}) is written as~\cite{PanossoMacedo:2019npm}
\begin{eqnarray}\label{Psi_scaling}
    \Psi^{(s)}(\tau,\sigma,\theta,\phi)=\Omega\Big[\Delta(\sigma)\Big]^{-s}\sum_{m=-\infty}^{+\infty}\cos^{\delta_1}(\theta/2)\sin^{\delta_2}(\theta/2)V_m(\tau,\sigma,\theta)\exp{(im\phi)}\, ,
\end{eqnarray}
where two exponents are $\delta_1=|m-s|$ and $\delta_2=|m+s|$, $\Delta(\sigma)=\Delta(r(\sigma))$, and $\Omega$ is the conformal factor, which is given by $\Omega=\sigma/\lambda$. The angle $\phi$-dependence is $e^{im\phi}$, where $m$ is the azimuthal number. With the substitution $x=\cos\theta$, the final regular form for the TE is achieved, 
\begin{eqnarray}\label{hyperboloidal_TE}
		0&=&\Bigg(\frac{\tilde{\Sigma}_0}{\beta}h_{,\sigma}\Big[2-\frac{\sigma^2\tilde{\Delta}h_{,\sigma}}{\tilde{\Sigma}_0\beta}\Big]-(1-x^2)[h_{,x}^2+\alpha^2]\Bigg)V_{m,\tau\tau}-[(1-x^2)V_{m,x}]_{,x}\nonumber\\
		&&+2i\alpha m\frac{\sigma^2}{\beta}V_{m,\sigma}+2\frac{\tilde{\Sigma}_0}{\beta}\Big[1-\frac{\sigma^2\tilde{\Delta}h_{,\sigma}}{\tilde{\Sigma}_0\beta}\Big]V_{m,\tau\sigma}-2(1-x^2)h_{,x}V_{m,\tau x}\nonumber\\
		&&+\Big[c_\tau-2i\alpha m\Big(1-\frac{\sigma^2h_{,\sigma}}{\beta}\Big)-[(1-x)\delta_1-(1+x)\delta_2]h_{,x}\Big]V_{m,\tau}\nonumber\\
		&&-\frac{\tilde{\Delta}^s}{\beta\sigma^{2s}}\Big[\frac{\sigma^{2(1+s)}\tilde{\Delta}^{1-s}}{\beta}V_{m,\sigma}\Big]_{,\sigma}+[(1+x)\delta_2-(1-x)\delta_1]V_{m,x}+\Big[\frac{2i\alpha m\sigma}{\beta}\nonumber\\
		&&-\frac{\sigma^{1-2s}\tilde{\Delta}^s}{\beta}\Big(\frac{\sigma^{2s}\tilde{\Delta}^{1-s}}{\beta}\Big)_{,\sigma}+2s-\Big(s-\frac{\delta_1+\delta_2}{2}\Big)\Big(1+s+\frac{\delta_1+\delta_2}{2}\Big)\Big]V_m\, ,
\end{eqnarray}
where expressions of functions $\tilde{\Sigma}_0$, $\tilde{\Delta}$, $c_\tau$ can be found in~\cite{PanossoMacedo:2019npm}, and the shift function $\beta(\sigma)=\rho(\sigma)-\sigma\rho^{\prime}(\sigma)$. In reality, the term proportional to $V_{m,\sigma\sigma}$ is $-\tilde{\Delta}\sigma^2/\beta^2$, which vanishes at $\sigma=0$ and $\sigma=\sigma_{+}$. This kind of behavior furnishes boundary conditions that assure the characteristics of the wave equation invariably direct outward from the numerical domain. Therefore, while seeking regular solutions, it is impermissible to impose any additional boundary conditions at the horizon or at future null infinity. Analogously to the behavior observed at the radial boundaries, the aforementioned equation undergoes degeneration at $x=\pm1$ as a result of the coefficient preceding $V_{m,xx}$ becoming zero. Similarly, since the corresponding regularity conditions must be imposed at the north and south poles of the spherical coordinate system, there is no necessity for additional boundary conditions. 

\begin{figure}
    \centering
    \includegraphics[width=0.3\linewidth]{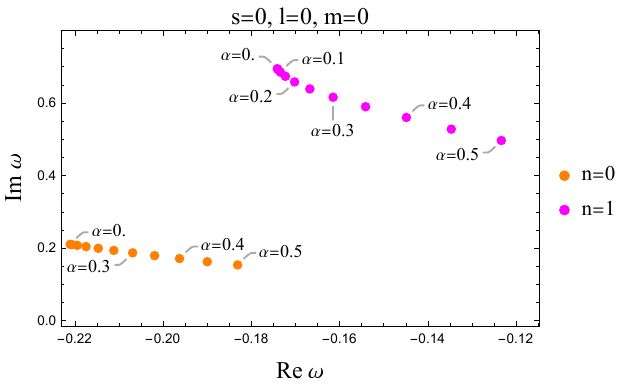}
    \hspace{0.1cm}
     \includegraphics[width=0.3\linewidth]{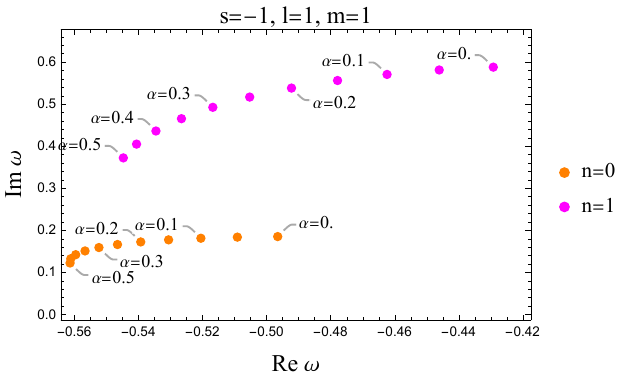}
     \hspace{0.1cm}
    \includegraphics[width=0.3\linewidth]{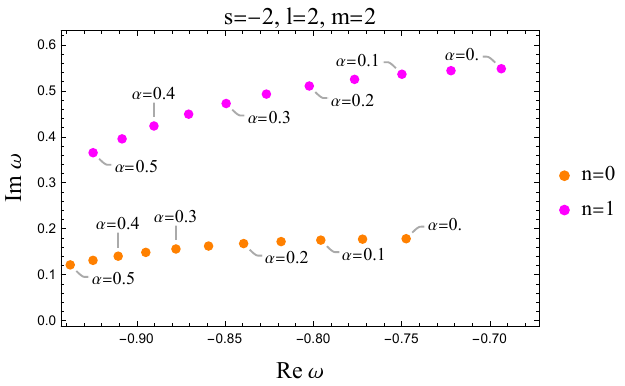}
    \includegraphics[width=0.3\linewidth]{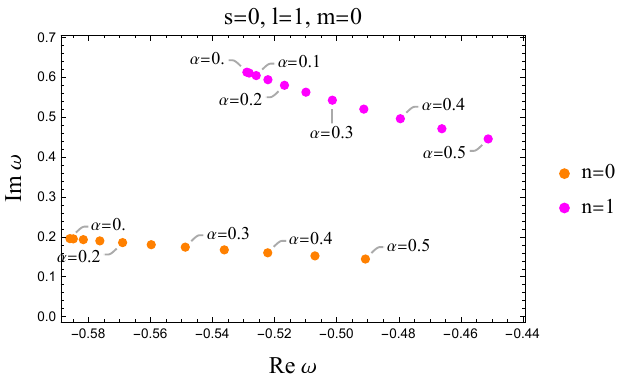}
     \hspace{0.1cm}
    \includegraphics[width=0.3\linewidth]{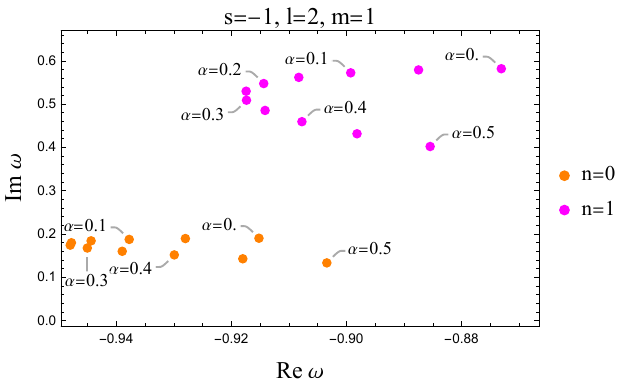}
     \hspace{0.1cm}
    \includegraphics[width=0.3\linewidth]{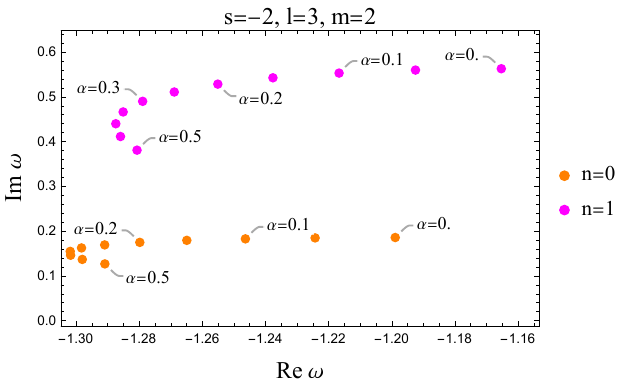}
    \caption{The QNM spectra of Kerr black holes. The top left panel is the QNM spectra for the modes with $s=0$, $l=0$, $m=0$, the top intermediate panel is the QNM spectra for the modes with $s=-1$, $l=1$, $m=1$, and the top right panel is the QNM spectra for the modes with $s=-2$, $l=2$, $m=2$. The bottom left panel is the QNM spectra for the modes with $s=0$, $l=1$, $m=0$, the bottom intermediate panel is the QNM spectra for the modes with $s=-1$, $l=2$, $m=1$, and the bottom right panel is the QNM spectra for the modes with $s=-2$, $l=3$, $m=2$. The magenta dots are the fundamental mode, while the orange dots are the first overtone. The corresponding $\alpha$ change between two adjacent points is $0.5$.}
    \label{fig:QNMs}
\end{figure}

If we apply the minimal gauge (radial function fixing gauge) and choose the length scale as $\lambda=r_{+}$, we have
\begin{eqnarray}\label{minimal_gauge_radial_function_fixing}
	\rho(\sigma)=1\, ,\quad \beta(\sigma)=1\, ,\quad h(\sigma,\theta)=-\frac{2}{\sigma}+4\mu\ln\sigma\, ,\quad \mu=\frac{1+\alpha^2}{2}\, ,
\end{eqnarray}
where the dimensionless mass $\mu$ and the dimensionless spin $\alpha$ are 
\begin{eqnarray}
    \mu=M/\lambda\, ,\quad \alpha=a/\lambda\, .
\end{eqnarray}
The reason why we choose the radial function fixing gauge is due to the computationtial simplicity without losing the correct physical results. At the same time, we perform a first-order reduction in time, i.e., $W_m=\partial_\tau V_m$, and the regular TE (\ref{hyperboloidal_TE}) in the minimal gauge (\ref{minimal_gauge_radial_function_fixing}) can be rewritten as two partial differential equations, involving first-order derivative respect to time and second-order derivative respect to $\sigma$ and $x$, i.e.,
\begin{eqnarray}\label{dynamics_TE}
	\partial_\tau u_m(\tau,\sigma,x)=iLu_m\, ,\quad u_{m}=
	\begin{bmatrix}
		V_m(\tau,\sigma,x)\\
		W_m(\tau,\sigma,x)
	\end{bmatrix}\, ,
\end{eqnarray}
where the operator $L$ is defined as
\begin{eqnarray}\label{operator_L}
	L=\frac{1}{i}
	\begin{bmatrix}
		0 & 1\\
		L_1 & L_2
	\end{bmatrix}\, .
\end{eqnarray}
We call $L$ is the time generator of the linear dynamics for the Kerr black hole. The expressions of the operators $L_1$ and $L_2$ can be found in the Appendix \ref{expressions_L1_and_L2}. Consider the Fourier transform of $u_m$ with respect to time $\tau$,
\begin{eqnarray}
    u_m(\tau, \sigma,x)\sim e^{i\omega\tau}u_m(\sigma,x)\, ,
\end{eqnarray}
we arrive at the two-dimensional eigenvalue problem as follows
\begin{eqnarray}\label{QNM_eigenvalue_problem}
    Lu_m(\sigma,x)=\omega u_m(\sigma,x)\, ,
\end{eqnarray}
in which boundary conditions are encoded in $L$. The eigenfunctions $u_m$ of the spectral problem above are QNMs, and they are regular in the region $(\sigma,x)\in[0,1]\times[-1,1]$. The QNM spectra of the Kerr black hole come from the above eigenvalue problem. We will use a two-dimensional pseudo-spectral method to solve the eigenvalue problem (see also~\cite{Xiong:2024urw} for the scalarized Kerr black holes). The details of such a method can be found in the Appendix \ref{two_dimensional_pseudo_spectral_method}, and we use the Chebyshev-Lobatto grid to obtain the QNM spectra. An important distinction between the spectra obtained from the two-dimensional eigenvalue problem and those from the one-dimensional problem (spherically symmetric) lies in the fact that the two-dimensional results encompass spectra of QNMs for all the angular momentum numbers $l\ge|m|$ and $l\ge|s|$. The method of identifying a specific mode is illustrated in~\cite{Leaver:1985ax} (see figure $2$ therein). For $\text{Re}\omega_{lmn}<0$, we extract the fundamental mode ($n=0$) and the first overtone ($n=1$) for each spins $s=-2$, $-1$, $0$, and depict them in Fig.\ref{fig:QNMs}, where the range of dimensionless spin $\alpha$ is $[0,0.5]$. For $\alpha=0.5$, the corresponding angular parameter $a$ is given by $0.8$, since the relation between $\alpha$ and $a$ is $a=2\alpha/(1+\alpha^2)$, where $M=1$. These results have been checked with the Berti data for the QNM spectra of the Kerr black hole~\cite{Berti:2009kk}, and we find that they are consistent. Considering the requirements of numerical accuracy, we use the resolution with $N=30$ for the modes with $s=0$, $l=m=0$, $n=1$ and $l=m+1=1$, $n=1$ and use $N=15$ for others.

\section{The construction of the norm}\label{norm}
In this section, we will construct the norm (or inner product) that is used among our numerical calculations. Similarly to the spherically symmetric case, we hope to construct the corresponding norm through the energy current integration method~\cite{Jaramillo:2020tuu,Gasperin:2021kfv}. It is known that in order to construct some norm, the complex conjugation of the field needs to use simultaneously. In terms of TE, the energy currents are constructed by two fields $\Psi^{(s)}$ with spin $s$ and $\Psi^{(-s)}$ with spin $-s$, whose detailed expressions for the energy current can be found in~\cite{Toth:2018qrx,Csukas:2019kcb}. However, it is known that the two fields are actually independent, so we have $\overline{\Psi^{(s)}}\neq\Psi^{(-s)}$ for $s\neq0$, where bar stands for complex conjugation. Especially, for $\overline{\Psi^{(s)}}|_{s=0}=\Psi^{(-s)}|_{s=0}$, the energy-momentum reduces to
\begin{eqnarray}\label{energy_current_s_0}
    T_{ab}=\frac{1}{2}\Big(\nabla_a\bar{\Phi}\nabla_b\Phi+\nabla_a\Phi\nabla_b\bar{\Phi}-g_{ab}g^{cd}\nabla_c\bar{\Phi}\nabla_d\Phi\Big)\, ,
\end{eqnarray}
where the notion $\Psi^{(0)}$ has been replaced by $\Phi$. Accordingly, when the spin is equal to be zero, i.e., $s=0$, Eq.(\ref{covariant_form_TE}) reduces to be the complex Klein-Gordon equation given by
\begin{eqnarray}\label{KG_equation}
    \nabla^a\nabla_a\Phi=0\, .
\end{eqnarray}
 
Based on the above description, we only focus on the spin $s=0$ case. Following Ref.\cite{Gasperin:2021kfv}, we will construct the norm from the definition of energy on a spatial slice $\Sigma_\tau$. In terms of the Killing vector field $(\partial_\tau)^a$, the energy current is defined as
\begin{eqnarray}\label{energy_current}
    E^a=T^a{}_b(\partial_\tau)^b=\frac{1}{2}\Big[\nabla^a\bar{\Phi}(\partial_\tau)^b\partial_b\Phi+\nabla^a\Phi(\partial_\tau)^b\partial_b\bar{\Phi}-(\partial_\tau)^ag^{cd}\nabla_c\bar{\Phi}\nabla_d\Phi\Big]\, ,
\end{eqnarray}
where the master function $\Phi$ from Eq.(\ref{Psi_scaling}) has the following expansion 
\begin{eqnarray}\label{Psi_scaling_s_0}
    \Phi=\Omega\sum_{m=-\infty}^{+\infty}\cos^{|m|}(\theta/2)\sin^{|m|}(\theta/2)V_m(\tau,\sigma,\theta)\exp{(im\phi)}\, .
\end{eqnarray}
The total energy defined on the hypersurface $\tau=\text{constant}$ is given by
\begin{eqnarray}\label{total_energy}
    {}^{[\tau]}E=\int_{\Sigma_\tau}n_aE^a\sqrt{h_{\tau}}\mathrm{d}\sigma\mathrm{d}\theta\mathrm{d}\phi\, ,
\end{eqnarray}
where the unit future directed normal of the hypersurface $\tau=\text{constant}$ is denoted by $n_a=(-\mathrm{d}\tau)_a/\sqrt{-g^{\tau\tau}}$, and $h_\tau$ is the determinant of the metric to the hypersurface $\tau=\text{constant}$ with $h_\tau$ is positive for our convention. In fact, to confirm the future directivity, one can compute 
\begin{eqnarray}
    n^\tau=n^a(\mathrm{d}\tau)_a=g^{ab}n_b(\mathrm{d}\tau)_a=-\frac{g^{ab}(\mathrm{d}\tau)_a(\mathrm{d}\tau)_b}{\sqrt{-g^{\tau\tau}}}=-\frac{g^{\tau\tau}}{\sqrt{-g^{\tau\tau}}}=\sqrt{-g^{\tau\tau}}>0\, .
\end{eqnarray}
Before performing the calculation, we need to have a certain foresight of the form of the result we will obtain. Eq.(\ref{total_energy}) presents a triple integral. However, since we are considering a two-dimensional eigenvalue problem, we retain the integrals along $\sigma$ and $x$ (or $\theta$) directions, and we need to integrate out the integral along $\phi$ direction. Substituting Eq.(\ref{energy_current}) and Eq.(\ref{Psi_scaling_s_0}) into Eq.(\ref{total_energy}) with $s=0$, we finally arrive at the total energy ${}^{[\tau]}E$ as follow
\begin{eqnarray}
    {}^{[\tau]}E&=&2\pi\sum_{m=-\infty}^{+\infty}\int_0^1\mathrm{d}\sigma\int_0^\pi\mathrm{d}\theta\Big[4^{-1-|m|}(\sin\theta)^{2|m|-1}\Big]\nonumber\\
    &&\times\Bigg\{\Big[2(\sigma-1)(\alpha^2 \sigma -1)\sin^2\theta+m^2\Big(\cos(2\theta )+3\Big)\Big]\bar{V}_mV_m\nonumber\\
    &&+\Big[2\sigma\sin^2\theta\Big(\alpha ^2 \sigma ^2-(\alpha ^2+1) \sigma -i \alpha  m \sigma +1\Big)\Big](\partial_\sigma\bar{V}_m)V_m+|m|\sin(2 \theta)(\partial_\theta\bar{V}_m)V_m\nonumber\\
    &&+\Big[2\sigma\sin^2\theta\Big(\alpha ^2 \sigma ^2-(\alpha ^2+1)\sigma+i\alpha  m \sigma +1\Big)\Big]\bar{V}_m(\partial_\sigma V_m)+|m|\sin(2\theta)\bar{V}_m(\partial_\theta V_m)\nonumber\\
    &&+\sin^2\theta\Big[-8 \alpha ^6 (\sigma -1) \sigma +\alpha ^4(-16 \sigma ^2+16 \sigma +8)+\alpha^2\cos(2\theta)\nonumber\\
    &&+\alpha ^2(-8 \sigma ^2+16 \sigma +15)+8 (\sigma +1)\Big](\partial_\tau\bar{V}_m)(\partial_\tau V_m)\nonumber\\
    &&+\Big[2 (\sigma -1) \sigma ^2(\alpha ^2 \sigma -1)\sin^2\theta\Big](\partial_\sigma\bar{V}_m)(\partial_\sigma V_m)+2\sin^2\theta(\partial_\theta\bar{V}_m)(\partial_\theta V_m)\Bigg\}\nonumber\\
    &\equiv&\sum_{m=-\infty}^{+\infty}E_{m}\, ,
\end{eqnarray}
where $E_{m}$ is called the mode energy and we have integrated out the coordinate $\phi$ by using $\int_0^{2\pi}e^{i(m-n)\phi}\mathrm{d}\phi=2\pi\delta_{m,n}$. Such name for $E_m$ is motivated by the case of spherically symmetric in~\cite{Gasperin:2021kfv}. By defining $x=\cos\theta$ and using some integration by parts, we find the mode energy $E_m$ is simplified into
\begin{eqnarray}\label{mode_energy}
    E_m&=&\frac{2\pi}{4^{1+|m|}}\int_0^1\mathrm{d}\sigma\int_{-1}^{1}\mathrm{d}x(1-x^2)^{|m|}\Bigg\{\Big[2\sigma(1+\alpha^2-2\alpha^2\sigma)+2|m|(1+|m|)\Big]\bar{V}_mV_m\nonumber\\
    &&+\Big[-8\alpha^6(\sigma-1) \sigma +\alpha ^4(-16 \sigma ^2+16 \sigma +8)+\alpha^2(2x^2-1)+\alpha ^2(-8 \sigma ^2+16 \sigma +15)+8 (\sigma +1)\Big]\bar{W}_mW_m\nonumber\\
    &&+\Big[2 (\sigma -1) \sigma ^2(\alpha ^2 \sigma -1)\Big](\partial_\sigma\bar{V}_m)(\partial_\sigma V_m)+2(1-x^2)(\partial_x\bar{V}_m)(\partial_x V_m)\Bigg\}\nonumber\\
    &&+\frac{2\pi}{4^{1+|m|}}\int_0^1\mathrm{d}\sigma\int_{-1}^{1}\mathrm{d}x2\alpha m\sigma^2(1-x^2)^{|m|}\Big[-i(\partial_\sigma\bar{V}_m)V_m+i\bar{V}_m(\partial_\sigma V_m)\Big]\, .
\end{eqnarray}
From the expression of the mode energy $E_m$ mentioned above, it is not difficult to see that the mode energy is always real and the first four terms of $E_m$ are positive ($\sigma\in[0,1]$ and $x\in[-1,1]$). Unfortunately, unlike the case of spherical symmetry, the mode energy of a Kerr black hole is not necessarily positive for the existence of last term of Eq.(\ref{mode_energy}). The total energy is defined as Eq.(\ref{total_energy}). This quantity can only be guaranteed to be conserved if the boundary is taken into account. Note that for the hyperboloidal coordinates, the contributions of the boundary exist. For example, one can refer to the study in~\cite{Csukas:2019kcb} (see Fig.5 therein). Hence, the non-positiveness of $E_m$ is possible to exist somehow. However, it is found that when $\alpha$ equals $0$, meaning the Kerr black hole returns to the Schwarzschild black hole, the resulting mode energy is positive. Additionally, as $m$ equals $0$, the mode energy is also positive no matter what $\alpha$ is. Considering that a fundamental requirement of a norm is its positive definiteness, and that it should preserve as much information about the original energy as possible,  we put forward a physically-motivated inner product denoted by $\langle\cdot,\cdot\rangle_E$ based on the mode energy (\ref{mode_energy}), i.e., 
\begin{eqnarray}\label{energy_norm_inner_product}
	&&\langle u_{m1},u_{m2}\rangle_E\nonumber\\
	&=&\frac{2\pi}{4^{1+|m|}}\int_0^1\mathrm{d}\sigma\int_{-1}^{1}\mathrm{d}x(1-x^2)^{|m|}\Bigg\{\Big[2\sigma(1+\alpha^2-2\alpha^2\sigma)+2|m|(1+|m|)\Big]\bar{V}_{m1}V_{m2}\nonumber\\
	&&+\Big[-8\alpha^6(\sigma-1) \sigma +\alpha^4(-16\sigma^2+16 \sigma +8)+\alpha^2(2x^2-1)+\alpha ^2(-8 \sigma ^2+16 \sigma +15)+8 (\sigma +1)\Big]\bar{W}_{m1}W_{m2}\nonumber\\
	&&+\Big[2 (\sigma -1) \sigma ^2(\alpha ^2 \sigma -1)\Big](\partial_\sigma\bar{V}_{m1})(\partial_\sigma V_{m2})+2(1-x^2)(\partial_x\bar{V}_{m1})(\partial_x V_{m2})\Bigg\}\, ,
\end{eqnarray}
where the corresponding norm is considered to be $\lVert u_{m}\rVert_E=\sqrt{\langle u_{m},u_{m}\rangle_E}$, and it is so-called the energy norm here. We see that $\partial_\sigma$ and $\partial_x$ are presented in Eq.(\ref{energy_norm_inner_product}), so the energy norm is actually an $H^1$-norm. This inner product (\ref{energy_norm_inner_product}) depends on $\alpha$ and $m$. Several kinds of definitions of the scalar product for QNMs have been considered in~\cite{Motohashi:2024fwt,London:2023aeo,London:2023idh,Green:2022htq}. However, there is a crucial difference between their scalar products and the inner product presented: our inner product does not depend on $\omega$, whereas their scalar product does depend on $\omega$. Furthermore, as $\langle u_m,u_m\rangle_E>0$, it can be said that we have defined a Hilbert space equiped with the norm $\lVert\cdot\rVert_E$. Since we will compare the convergence of resolvents under different norms in subsequent sections, we present some norms that will be used. The standard $L^2$-norm of $u_m$ is given by
\begin{eqnarray}\label{L2_norm}
    \lVert u_m\rVert_{L^2}^2=\int_{0}^1\mathrm{d}\sigma\int_{-1}^1\mathrm{d}x\Big(|V_m|^2+|W_m|^2\Big)\, .
\end{eqnarray}
The inner product associated with above $L^2$-norm is defined naturally. For higher-order $H^k$ Sobolev norms with $k\ge2$, we add a term~\cite{Boyanov:2023qqf,Besson:2024adi}
\begin{eqnarray}\label{Hk_norm}
    (\partial_\sigma^k\bar{V}_m)(\partial_\sigma^kV_m)+(\partial_x^k\bar{V}_m)(\partial_x^k V_m)+(\partial_\sigma^k\bar{W}_m)(\partial_\sigma^kW_m)+(\partial_x^k\bar{W}_m)(\partial_x^k W_m)\, ,
\end{eqnarray}
into the original energy norm, where we ignore the mixed derivative terms for simplicity. 

In order to calculate the pseudospectrum of the operator $L$ explicitly, we should use the discretized versions of the inner products associated with the corresponding norms. Using the so-called Gram matrix, the continuous version of the inner product can be translated into the matrix quadratic form, i.e.,
\begin{eqnarray}
	\langle \mathbf{u}_{m1},\mathbf{u}_{m2}\rangle=\begin{bmatrix}
		\mathbf{V}^{\star}_{m1} & \mathbf{W}^{\star}_{m1}
	\end{bmatrix}\cdot
	\mathbf{G}\cdot
	\begin{bmatrix}
		\mathbf{V}_{m2}\\
		\mathbf{W}_{m2}
	\end{bmatrix}\, ,
\end{eqnarray}
where the symbol $\star$ stands for the Hermite conjugate, and $\cdot$ is the multiplication of the matrix. For convenience, we denote the Gram matrices of the energy norm ($H^1$-norm), $L^2$-norm and $H^k$-norm as $\mathbf{G}_E$, $\mathbf{G}_{L^2}$ and $\mathbf{G}_{H^k}$, respectively. In Appendix \ref{Gram_matrix}, we explicitly write the forms of $\mathbf{G}_E$ and $\mathbf{G}_{L^2}$. Furthermore, from Eq.(\ref{Hk_norm}), we have the relationship between $\mathbf{G}_E$ and $\mathbf{G}_{H^k}$ which is given by
\begin{eqnarray}
    \mathbf{G}_{H^k}=\mathbf{G}_E+
    \begin{bmatrix}
        \Delta \mathbf{G} & \mathbf{0}\\
        \mathbf{0} & \Delta \mathbf{G}
    \end{bmatrix}\, ,
\end{eqnarray}
where $\mathbf{0}$ represents the zero matrix and $\Delta \mathbf{G}$ is given by
\begin{eqnarray}\label{Delta_G}
    \Delta \mathbf{G}=(\mathbf{I}\otimes(\mathbf{D}_\sigma)^k)^T\cdot\mathbb{W}\cdot(\mathbf{I}\otimes(\mathbf{D}_\sigma)^k)+((\mathbf{D}_x)^k\otimes\mathbf{I})^T\cdot\mathbb{W}\cdot((\mathbf{D}_x)^k\otimes\mathbf{I})\, .
\end{eqnarray}
In the above equation, $\mathbb{W}$ is the two-dimensional weight matrix which is a diagonal matrix, and $\mathbf{D}_\sigma$ and $\mathbf{D}_x$ are the differentiation matrices for some Chebyshev grid. In actual practice, the computation of pseudospectra involves the Cholesky decomposition of the Gram matrix, which necessitates that the Gram matrix be strictly positive definite. Due to the presence of the term $(1-x^2)^{|m|}$ in the energy inner product, we observe that when the Chebyshev-Lobatto grid is used, the resulting Gram matrix $\mathbf{G}_E$ is not strictly positive definite (positive semi-definite), indicating potential numerical difficulties. To avoid this issue, we opt to discretize the inner product using the Chebyshev-Gauss grid, under which the resulting Gram matrix $\mathbf{G}_E$ is indeed strictly positive definite. 

\section{The pseudospectrum and the numerical convergence test for the norm of resolvent}\label{pseudospectrum_and_convergence}
In the previous sections, we not only obtain
the QNM eigenvalue problem (\ref{QNM_eigenvalue_problem}) in the hyperboloidal framework (see Sec.\ref{TE_QNMs}) but also some norms (see Sec.\ref{norm}), which will be used. Therefore, in this section, we will study the pseudospectrum of the Kerr black hole with spin $s=0$ case. First, the definition of the pseudospectrum for the operator $L$ is supposed to be given. Given $\epsilon>0$ and some norm $\lVert\cdot\rVert$, the $\epsilon$-pseudospectrum $\sigma_{\epsilon}(L)$ is defined as~\cite{trefethen2020spectra}
\begin{eqnarray}\label{pseudospectrum_definition}
	\sigma_{\epsilon}(L)=\{\omega\in\mathbb{C}:\lVert R_{L}(\omega)\rVert=\lVert(L-\omega\mathbb{I})^{-1}\rVert>1/\epsilon\}\, ,
\end{eqnarray}
where $R_{L}(\omega)$ is called the resolvent operator. In the limit $\epsilon\to0$, the set $\sigma_\epsilon(L)$ reduces to the spectrum set $\sigma(L)$, whose elements are the spectrum $\omega_n$. The quantity $\epsilon$ serves as a measure of the ``proximity" between points in $\sigma_\epsilon(L)$ and the spectrum $\omega_n$, offering a clear interpretation of perturbations to the underlying operator. Therefore, the shape and size of the $\epsilon$-pseudospectrum regions quantify the spectrum (in)stability of the operator $L$. Note that for our current study, the expression of $L$ can be found in Appendix \ref{expressions_L1_and_L2}, where $s=0$ should be substituted in $L$. Numerical methods for computing pseudospectra can be found in numerous references~\cite{Jaramillo:2020tuu,Cao:2024oud,Chen:2024mon,Besson:2024adi,trefethen2020spectra,trefethen_1999}, and the specific details will not be described here. The crucial step in the calculation is to convert the norm of the operator in Eq.(\ref{pseudospectrum_definition}) into the $2$-norm of a matrix (see Chapter 43 in~\cite{trefethen2020spectra}). Note that the $2$-norm of a matrix is equal to its maximum singular value, and the $2$-norm of the inverse of a matrix corresponds to the reciprocal of its minimum singular value.

\begin{figure}[htbp]
	\centering
    \includegraphics[width=0.4\textwidth]{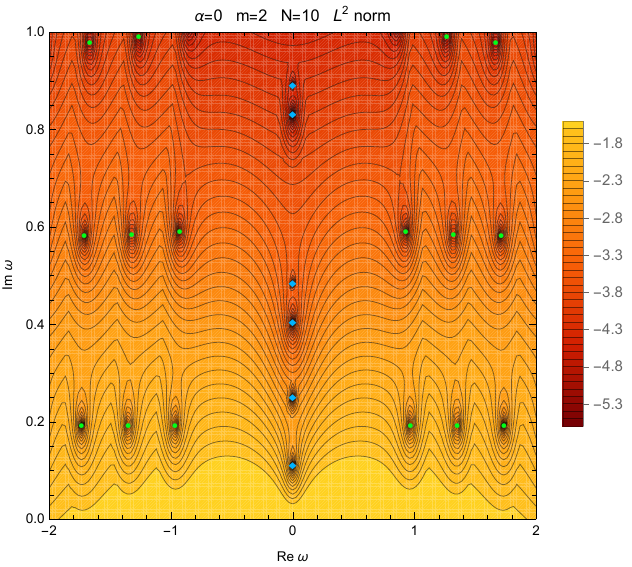}
    \hspace{0.4cm}
	\includegraphics[width=0.4\textwidth]
    {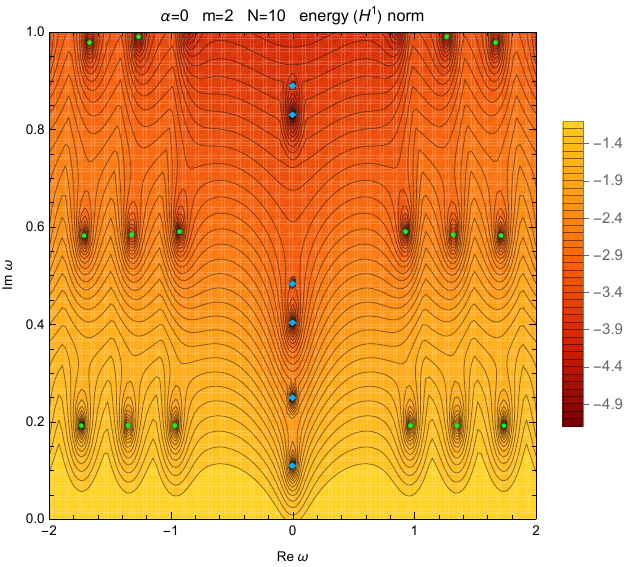}
    \includegraphics[width=0.4\textwidth]
    {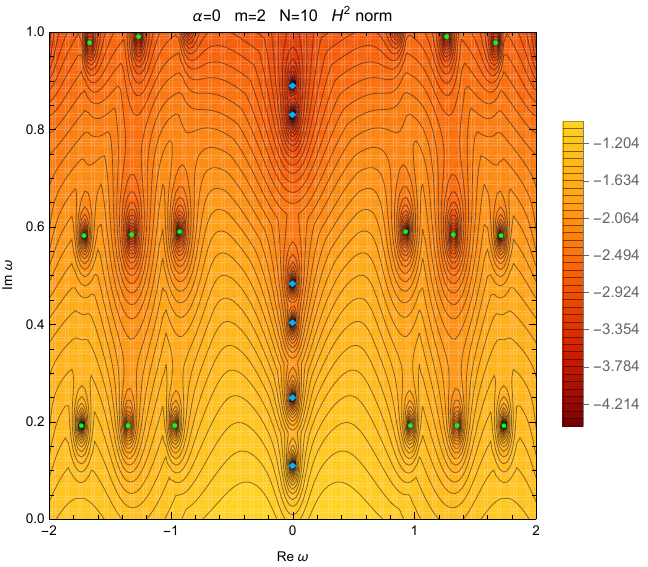}
    \hspace{0.4cm}
    \includegraphics[width=0.4\textwidth]{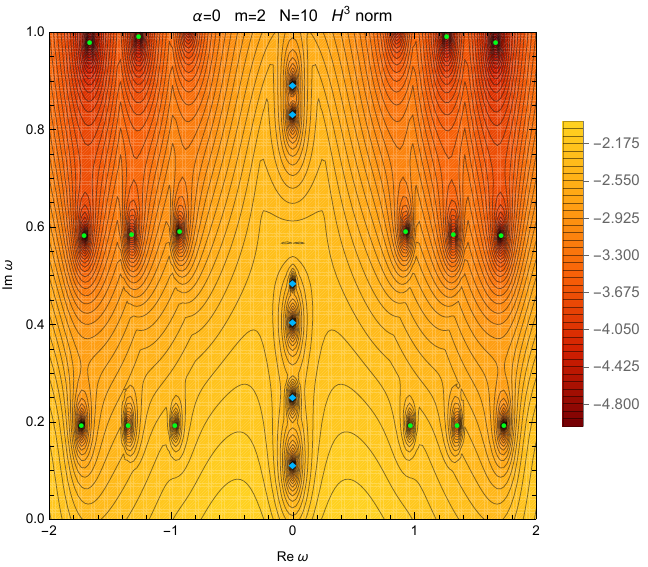}
    \includegraphics[width=0.4\textwidth]{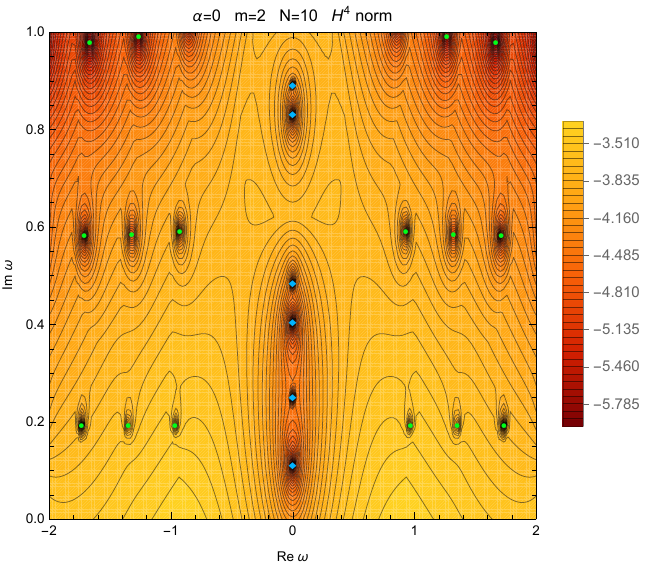}
    \hspace{0.4cm}
    \includegraphics[width=0.4\textwidth]{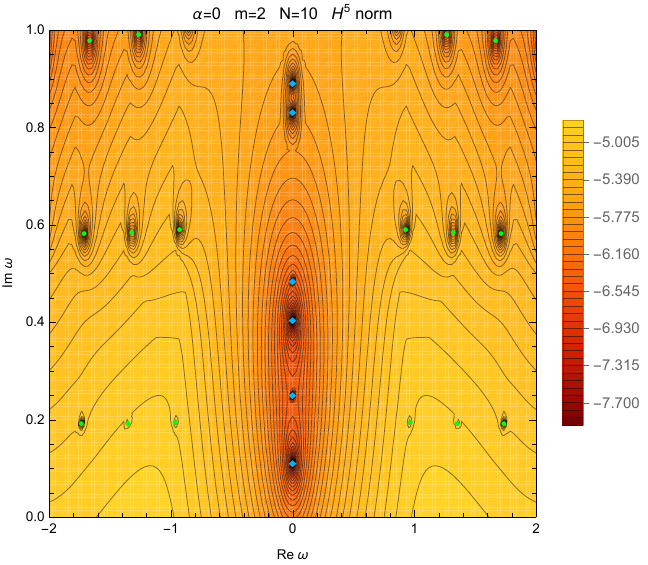}
	\caption{The pseudospectra $\log_{10}[\sigma_{\epsilon}(L)]$ for $L^2$-norm, energy ($H^1$) norm, $H^2$-norm, $H^3$-norm, $H^4$-norm and $H^5$-norm of Kerr black hole with $\alpha=0$ and $m=2$. Here, the resolution for the operator $L$ is given by $N=10$. For all panels, the scopes of drawings are all limited to $\text{Re}\omega_{\text{max}}=2$, $\text{Re}\omega_{\text{min}}=-2$, $\text{Im}\omega_{\text{min}}=0$ and $\text{Im}\omega_{\text{max}}=1$. The resolutions of pseudospectra figures are all set as $\Delta\text{Re}\omega=4/150$ and $\Delta\text{Im}\omega=1/150$. In addition, the number of contour lines is $40$ for these six panels.}
    \label{pseudospectra_alpha_0_m_2}
\end{figure}

In Fig.\ref{pseudospectra_alpha_0_m_2}, we present the results of the pseudospectra for different norms containing $L^2$-norm, energy ($H^1$) norm, $H^2$-norm, $H^3$-norm, $H^4$-norm and $H^5$-norm within $\alpha=0$ and $m=2$. For more pseudospectra of other $\alpha$'s, one can refer to Appendix \ref{other_alpha}, where we use $\alpha=0.25$ to get the pseudospectra. An important distinction between the pseudospectrum when $\alpha=0$ and when $\alpha\neq0$ lies in the fact that the pseudospectrum contour figures are no longer symmetric on the imaginary axis. This is also a direct consequence of the QNM spectrum no longer being symmetric about the imaginary axis for $\alpha\neq0$. For each panel in Fig.\ref{pseudospectra_alpha_0_m_2}, the green and blue points represent the eigenvalues of the matrix operator $\mathbf{L}$ with resolution $N=10$ in the complex plane. Note that such a resolution is enough to solve the QNM spectra for low-order modes. The family of blue diamond points is located on the imaginary axis, and these points are introduced due to the discretization process and are considered to be continuous distributions when the original operator $L$ is considered. The green circle points represent the QNM spectra of the Kerr black hole. The three blue dotts in the bottom right corner of Fig.\ref{pseudospectra_alpha_0_m_2} are the fundamental modes with $l=2$, $l=3$, $l=4$, respectively. The three blue dotts in the intermediate right of Fig.\ref{pseudospectra_alpha_0_m_2} are the first overtones with $l=2$, $l=3$, $l=4$, respectively. As the overtone number $n$ increases, the stabilities of the spectrum $\omega_{n}$ become worse. Fig.\ref{pseudospectra_alpha_0_m_2} illustrates that spectrum instability also exists for Kerr black holes as open sets are formed by the contour lines of the $\epsilon$-pseudospectrum. This also demonstrates that the phenomenon of spectrum instability is ubiquitous in the Kerr black hole. This phenomenon is not surprising, as the QNM system of black holes constitutes a non-Hermitian system, which naturally exhibits spectrum instability~\cite{Ashida:2020dkc}.

\begin{figure}[htbp]
	\centering
    \includegraphics[width=0.4\textwidth]{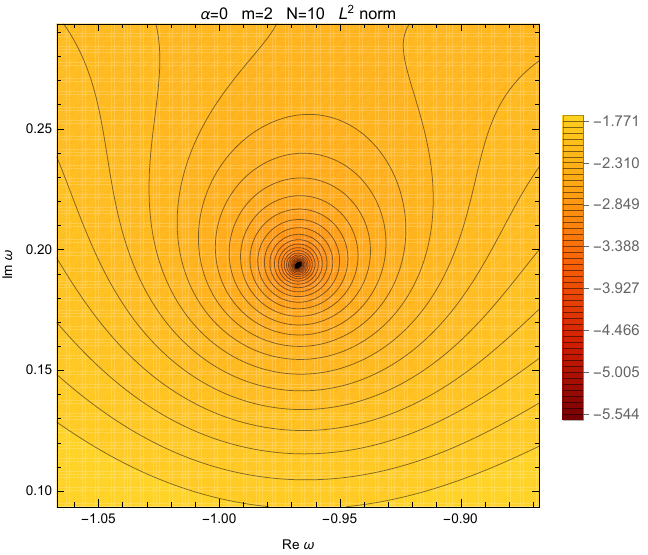}
    \hspace{0.4cm}
	\includegraphics[width=0.4\textwidth]
    {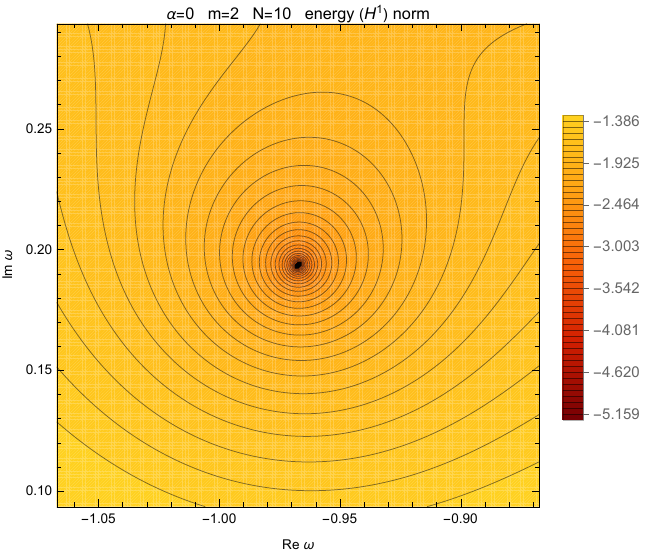}
    \includegraphics[width=0.4\textwidth]
    {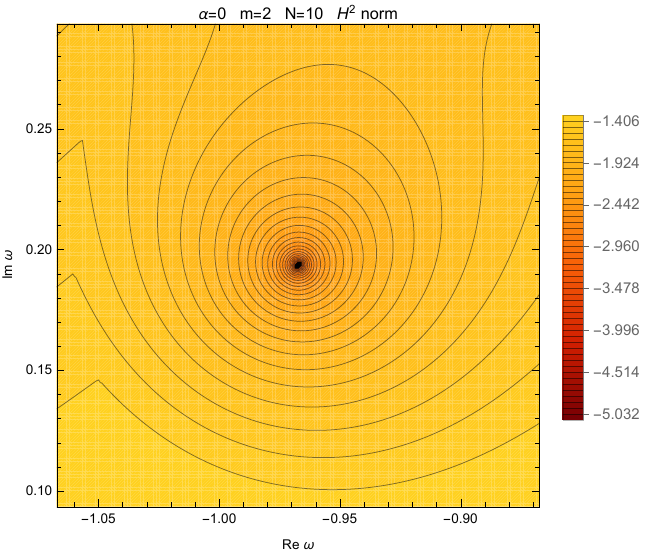}
    \hspace{0.4cm}
    \includegraphics[width=0.4\textwidth]{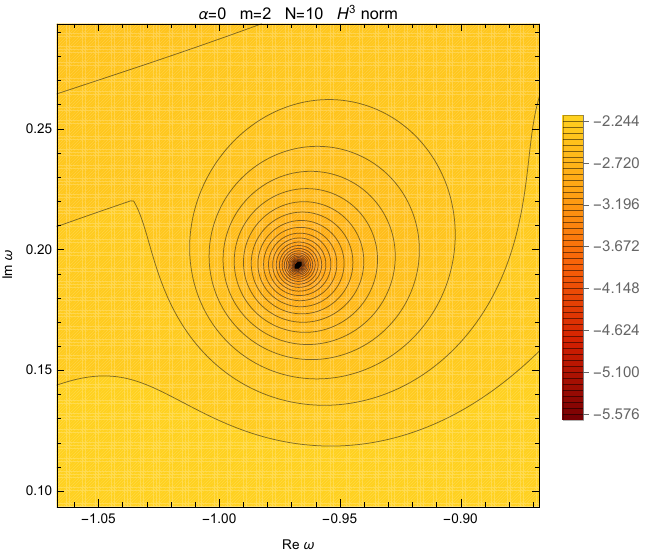}
    \includegraphics[width=0.4\textwidth]{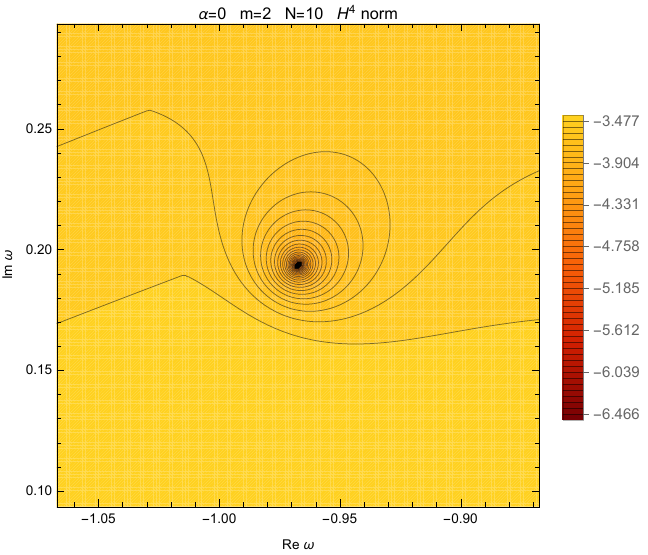}
    \hspace{0.4cm}
    \includegraphics[width=0.4\textwidth]{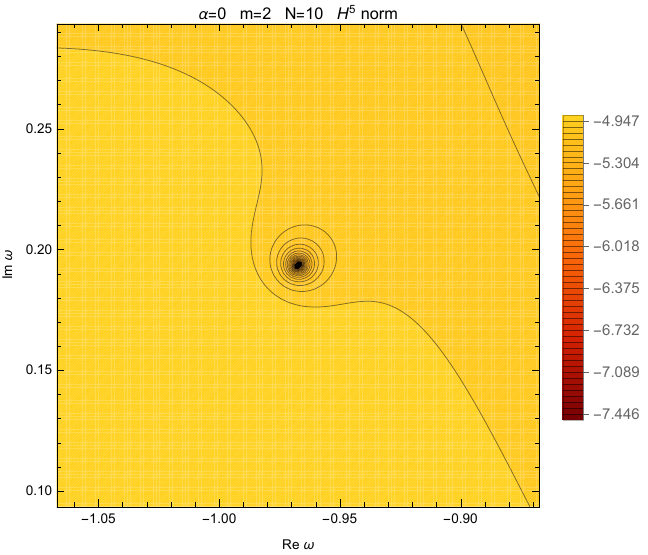}
	\caption{The pseudospectra $\log_{10}[\sigma_{\epsilon}(L)]$ for $L^2$-norm, energy ($H^1$) norm, $H^2$-norm, $H^3$-norm, $H^4$-norm and $H^5$-norm of Kerr black hole with $\alpha=0$ and $m=2$ around the mode with $l=2$ and $n=0$. For all panels, the scopes of drawings are all limited to $\text{Re}\omega_{\text{max}}=\text{Re}\omega_{l=2,n=0}+0.1$, $\text{Re}\omega_{\text{min}}=\text{Re}\omega_{l=2,n=0}-0.1$, $\text{Im}\omega_{\text{min}}=\text{Im}\omega_{l=2,n=0}-0.1$ and $\text{Im}\omega_{\text{max}}=\text{Im}\omega_{l=2,n=0}+0.1$. The resolutions of figures are all set as $\Delta\text{Re}\omega=\Delta\text{Im}\omega=1/750$. In addition, the number of contour lines is $50$ for all panels.}
    \label{pseudospectra_alpha_0_m_2_l_2_n_0}
\end{figure}

From Fig.\ref{pseudospectra_alpha_0_m_2}, it can be seen that the shapes of the pseudospectra depend on the choice of different norms. It is worth noting that adjacent contour lines in each figure have the same $\epsilon$ interval. It is observed that as $k$ increases, the contour lines located farther away from the QNM spectra become progressively sparser, whereas those located closer to the QNM spectra exhibit a marked tendency to become denser. Especially for the mode with $l=2$ and $n=0$, this phenomenon is particularly prominent. When $k$ is relatively large, the norm of the resolvent changes very slowly in regions slightly away from this mode, whereas it grows very rapidly as one approaches this mode. Therefore, we further zoom in such mode in Fig.\ref{pseudospectra_alpha_0_m_2_l_2_n_0}. In these subfigures, we have increased the number of contour lines to $50$, where the plot region is a square area with a side length of $0.2$ centered on the mode $l=2$, $m=2$, and $n=0$. It is observed that as $k$ increases, closed circles will encompass an increasingly smaller area, while the contour lines of the remaining regions become increasingly sparse.

\begin{figure}[htbp]
	\centering
    \includegraphics[width=0.33\textwidth]{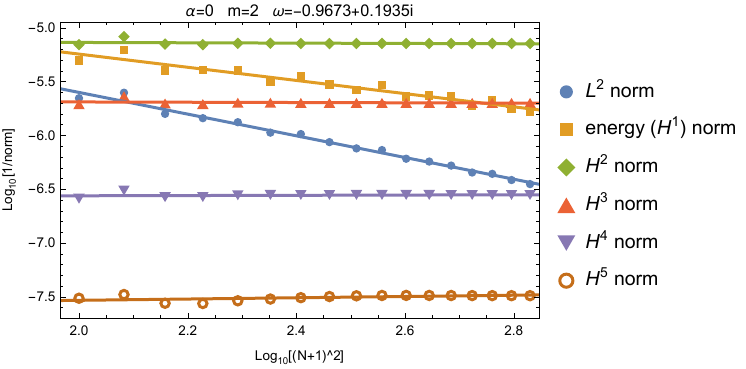}
    \includegraphics[width=0.33\textwidth]{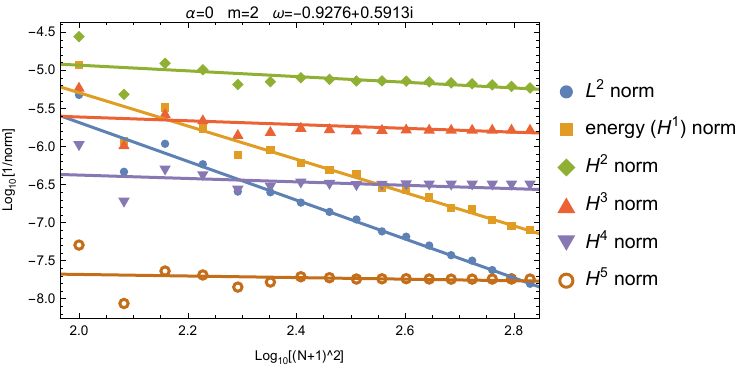}
    \includegraphics[width=0.33\textwidth]{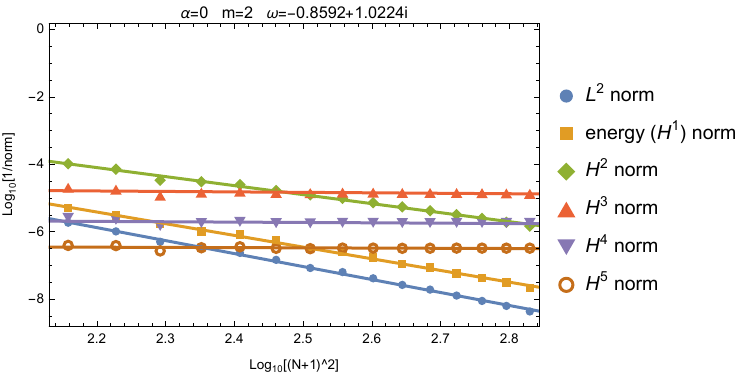}
    \includegraphics[width=0.33\textwidth]{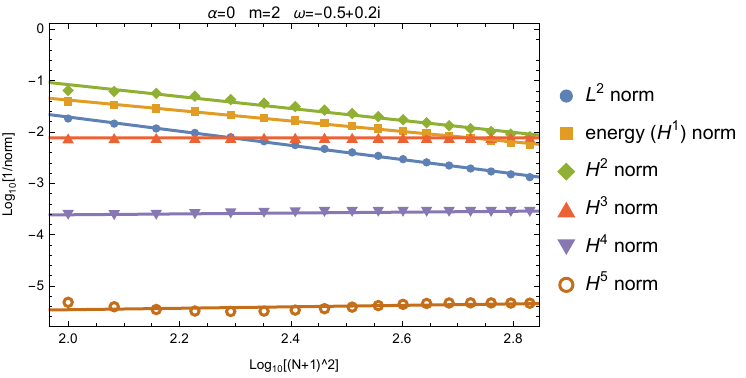}
    \includegraphics[width=0.33\textwidth]{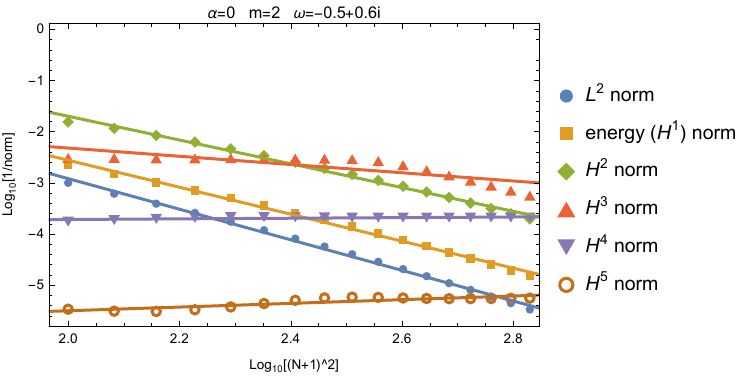}
    \includegraphics[width=0.33\textwidth]{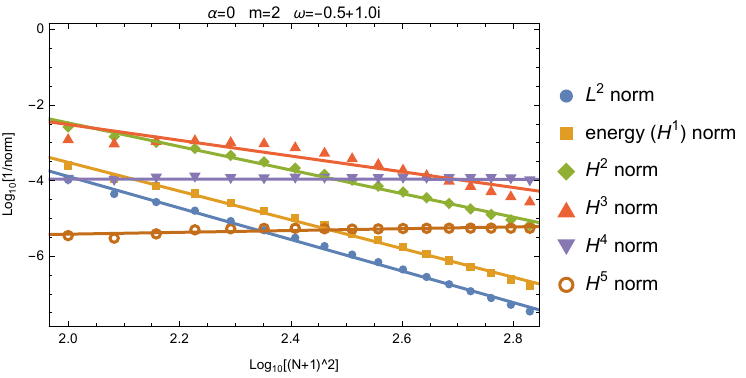}
	\caption{The inverse of the norm of $R_{\mathbf{L}}(\omega)$ as function of the numerical resolution $N$ (from $N=9$ to $N=25$), where the horizontal and vertical coordinates are $\log_{10}[(N+1)^2]$ and $\log_{10}(1/\lVert R_{\mathbf{L}}(\omega)\rVert)$, respectively. Different colors represent different norms. We use the least-squares method to fit the results using a linear model. The top panels for the complex numbers are in the vicinity of the QNM spectra, where the QNM spectra ($l=2$) are the modes with $n=0$, $n=1$, $n=2$. The bottom panels for the complex numbers are away from the QNM spectra.}
    \label{convergence_test_alpha_0_m_2}
\end{figure}

In the above, we present the traits of the pseudospectrum of Kerr black hole, but we can only compute it through numerical methods. This means that we only consider the finite rank approximant of the operator $L$. This prompts us to consider the convergence of the norm of the resolvent with respect to resolution $N$. Similar to the Ref.\cite{Boyanov:2023qqf} studying convergence of pseudospectrum in the AdS case, we characterize the convergence of the pseudospectrum of Kerr black hole as follows. First, we choose some complex number $\omega$ in complex plane. Second, we consider a finite rank approximant $\mathbf{L}$ (by using the Chebyshev collocation method) of the operator $L$ and calculate the norm of its resolvent $\lVert R_{\mathbf{L}}(\omega)\rVert=\lVert(\mathbf{L}-\omega \mathbf{I})^{-1}\rVert$, where the norms are chosen as $L^2$-norm, energy norm and $H^k$-norm. Third, we take the limit of $\lVert R_{\mathbf{L}}(\omega)\rVert$ as the resolution $N\to\infty$. Here, we show the convergence properties with the parameter set by $\alpha=0$ and $m=2$ in Fig.\ref{convergence_test_alpha_0_m_2}, in which three complex numbers are close to the QNM spectra and the other three complex numbers are away from the QNM spectra. Note that the number of grid points is $(N+1)^2$, and $N$ ranges from $N=9$ to $N=25$, here. We utilized the least-squares method to perform a linear fit based on the results and found that the fitting results are satisfactory. Consequently, we conclude that for a given $\omega$, the relationship between $\log_{10}(1/\lVert R_{\mathbf{L}}(\omega)\rVert)$ and $\log_{10}[(N+1)^2]$ can be roughly approximated as linear. We find that the convergence of the resolvent norm exhibits the following characteristics among the norm defined above and the complex number $\omega$. For each panel of Fig.\ref{convergence_test_alpha_0_m_2}, as $k$ increases, the convergence of the norm of $ R_{\mathbf{L}}(\omega)$ improves, regardless of whether $\omega$ is near the QNM spectrum or not. Through a horizontal comparison of Fig.\ref{convergence_test_alpha_0_m_2}, it can be seen that for the same $k$, as the imaginary part of $\omega$ increases, the convergence of the norm of the resolvent at that point will become increasingly poorer. Furthermore, through a vertical comparison of Fig.\ref{convergence_test_alpha_0_m_2}, we also find that the convergence of the norm at points farther away from the QNM spectra is worse than that at points closer to the QNM spectra.

%======================================%
%<<<<<<<<<<< Conclusions >>>>>>>>>>>>>%
%======================================%
\section{Conclusions and discussion}\label{sec: conclusions}
In this paper, we investigate the QNMs and the pseudospectrum of the Kerr black hole using the hyperboloidal framework~\cite{PanossoMacedo:2019npm}. After a reduction of order in time, we obtain hyperbolic partial differential equations of first order in time and second order in two-dimensional space $(\sigma,x)$. The advantage of such a scheme is that the QNM boundary conditions are built into the ``bulk" of the operator $L$, and no additional boundary conditions are required. The light cones point outward at the boundary of the computation domain, simplifying the boundary conditions to only requiring a regular solution, which is trivially satisfied in numerical calculations~\cite{PanossoMacedo:2024nkw}. The QNM spectrum is derived from the eigenvalue problem (\ref{QNM_eigenvalue_problem}), where the numerical method used is the Chebyshev pseudo-spectral method based on a tensor product grid (see Fig.\ref{tensor_product_grid}). To demonstrate the reliability of our method, we compare our results with those obtained by Berti and find them to be consistent.

Motivated by approaches to the construction of physical norms under spherical symmetry in~\cite{Jaramillo:2020tuu,Gasperin:2021kfv}, we construct the inner product (\ref{energy_norm_inner_product}) according to the mode energy (\ref{mode_energy}) in the Kerr black hole. This inner product is a double integral in the region $(\sigma,x)\in[0,1]\times[-1,1]$. Note that the mode energy (\ref{mode_energy}) is not always strictly positive. Of course, when the mode energy returns to the spherically symmetric scenario ($\alpha=0$), it remains strictly positive, where the positive definiteness of the norm is an important prerequisite for pseudospectra. So far, in the calculations of the pseudospectra of QNMs in the context of gravitational physics, we have not been able to analyze the pseudospectrum of an operator from an analytical perspective as in non-Hermitian quantum mechanics~\cite{Krejcirik:2014kaa}. Instead, we can only use numerical methods to approximate the operator $L$ with finite rank. Therefore, it is necessary to consider the convergence of the pseudospectrum, which refers to the change in the norm of the resolvent as the resolution varies. Given this, in addition to the original energy norm, we incorporate an additional norm term (see Eq.(\ref{Hk_norm})) to obtain an $H^k$-norm for investigating the convergence. In the main text, we display the pseudospectra of different norms with the parameters $\alpha=0$, $m=2$ in Fig.\ref{pseudospectra_alpha_0_m_2}, and the case with $\alpha=0.25$, $m=2$ is shown in the Appendix \ref{other_alpha}. For the convergence aspect, we find that as $k$ increases, the convergence of the norm of $R_{\mathbf{L}}(\omega)$ improves in terms of fixed $\omega$. Simultaneously, an augmentation in the imaginary component of $\omega$ may impair the convergence of the pseudospectrum, given the same norms.

Lastly, we provide an annotation on the pseudospectrum in which $s$ is not equal to $0$. The scenario where $s\neq0$ just means that the norm cannot be constructed using the energy current method. However, this does not imply that the pseudospectrum for this scenario cannot be computed. In Eq.(\ref{dynamics_TE}) and Eq.(\ref{operator_L}), we have already regularized the QNM functions. Consequently, given certain norms, the pseudospectrum for $s\neq0$ can be calculated in a similar way.

There are some things that can be further explored within the background of Kerr. Here, we list some investigations that can be done in the future.
\begin{enumerate}
    \item In general relativity, the hyperboloidal framework of the Kerr black hole has been worked out~\cite{PanossoMacedo:2019npm}. The hyperboloidal method provides us with a way to calculate the QNM frequencies by computing the eigenvalues of matrices. However, hyperboloidal coordinates in some modified theories of gravity, such as the Einstein-scalar-Gauss-Bonnet (EsGB) gravity theory, have not been studied yet. For modified theories of gravity, it is attractive for us to find a corresponding time generator operator similar to $L$ in Eq.(\ref{operator_L}), where this operator has absorbed the boundary conditions of QNMs.
    
    \item Motivated by the recent work~\cite{Besson:2024adi}, we will study a hyperboloidal Keldysh's approach on the Kerr black hole in order to achieve the QNM expansions of black hole perturbations. As a numerical study, the foremost task is the calculation of time-domain waveforms in the Kerr case. Therefore, this series of references~\cite{markakis2019timesymmetry,OBoyle:2022yhp,Markakis:2023pfh,DaSilva:2023xif,DaSilva:2024yea} is extremely helpful for us, since we have derived the dynamics of TE [see Eq.(\ref{dynamics_TE})-Eq.(\ref{operator_L})]. 
\end{enumerate}
Furthermore, considering that gravitational waves radiate from infinity or are absorbed by black holes, the perturbation systems of black holes are generally non-Hermitian. Non-Hermitian systems are extensively studied in other fields of physics~\cite{Ashida:2020dkc,Krejcirik:2014kaa}. Therefore, can we apply some research methods or techniques from other fields to the problem of QNMs of black holes? This is a fascinating topic.

%======================================%
%<<<<<<<<<< Acknowledgement >>>>>>>>>>>%
%======================================%
\section*{Acknowledgement}
We are grateful to Long-Yue Li, Xia-Yuan Liu, and Libo Xie for helpful discussions. This work is supported in part by the National Key R\&D Program of China Grant No.2022YFC2204603, by the National Natural Science Foundation of China with grants No.12475063, No.12075232 and No.12247103. This work is also supported by the National Key Research and Development Program of China Grant No.2020YFC2201501, by the National Natural Science Foundation of China Grants No.12475067 and No.12235019. This work is also supported by the National Natural Science Foundation of China with grants No.12235019 and No.11821505.

\appendix
\section{The expressions of the operator $L_1$ and $L_2$}\label{expressions_L1_and_L2}
In this appendix, we give the explicit expression of the operator $L$ in Eq.(\ref{operator_L}). The expressions of the operators $L_1$ and  $L_2$ are written as
\begin{eqnarray}
	L_1={}^{[x]}L_1^{2}(\sigma,x)\frac{\partial^2}{\partial x^2}+{}^{[\sigma]}L_1^{2}(\sigma,x)\frac{\partial^2}{\partial \sigma^2}+{}^{[x]}L_1^{1}(\sigma,x)\frac{\partial}{\partial x}+{}^{[\sigma]}L_1^{1}(\sigma,x)\frac{\partial}{\partial \sigma}+L_1^0(\sigma,x)\, ,
\end{eqnarray}
\begin{eqnarray}
	L_2={}^{[\sigma]}L_2^{1}(\sigma,x)\frac{\partial}{\partial \sigma}+L_2^0(\sigma,x)\, ,
\end{eqnarray}
where seven functions ${}^{[x]}L_1^{2}(\sigma,x)$, ${}^{[\sigma]}L_1^{2}(\sigma,x)$, ${}^{[x]}L_1^{1}(\sigma,x)$, ${}^{[\sigma]}L_1^{1}(\sigma,x)$, $L_1^0(\sigma,x)$, ${}^{[\sigma]}L_2^{1}(\sigma,x)$, and $L_2^0(\sigma,x)$ associated with $(\sigma,x)$ are
\begin{eqnarray}
	{}^{[x]}L_1^{2}(\sigma,x)=\frac{1-x^2}{4(\alpha ^2+1)[\alpha ^2 (1-\sigma )+1] [(\alpha^2+1) \sigma +1]-\alpha ^2(1-x^2)}\, ,
\end{eqnarray}
\begin{eqnarray}
	{}^{[\sigma]}L_1^{2}(\sigma,x)=\frac{(1-\sigma ) \sigma ^2(1-\alpha ^2 \sigma)}{4(\alpha ^2+1)[\alpha^2 (1-\sigma)+1][(\alpha ^2+1) \sigma +1]-\alpha ^2(1-x^2)}\, ,
\end{eqnarray}
\begin{eqnarray}
	{}^{[x]}L_1^{1}(\sigma,x)=\frac{\delta_1 (1-x)-\delta_2(x+1)-2 x}{4(\alpha ^2+1)[\alpha ^2 (1-\sigma )+1][(\alpha ^2+1) \sigma +1]-\alpha ^2(1-x^2)}\, ,
\end{eqnarray}
\begin{eqnarray}
	{}^{[\sigma]}L_1^{1}(\sigma,x)=\frac{\sigma\Big(2 (s+1)-\sigma[-4 \alpha ^2 \sigma +2 i \alpha  m+(\alpha ^2+1) (s+3)]\Big)}{4(\alpha ^2+1)[\alpha ^2 (1-\sigma)+1][(\alpha^2+1) \sigma +1]-\alpha ^2(1-x^2)}\, ,
\end{eqnarray}
\begin{eqnarray}
	L_1^0(\sigma,x)=\frac{-\sigma [\alpha^2(1-2\sigma )+1]-2 i \alpha m\sigma -(\alpha ^2+1) s \sigma +\Big(\frac{\delta_1+\delta_2}{2}+s+1\Big) \Big(-\frac{\delta_1+\delta_2}{2}+s\Big)}{4(\alpha ^2+1)[\alpha ^2 (1-\sigma )+1][(\alpha ^2+1) \sigma +1]-\alpha ^2(1-x^2)}\, ,
\end{eqnarray}
\begin{eqnarray}
	{}^{[\sigma]}L_2^{1}(\sigma,x)=\frac{2\Big(\alpha ^2\sigma^2-2(\alpha ^2+1) \sigma ^2[\alpha^2 (1-\sigma)+1]+1\Big)}{4(\alpha ^2+1)[\alpha ^2 (1-\sigma )+1] [(\alpha ^2+1) \sigma +1]-\alpha ^2(1-x^2)}\, ,
\end{eqnarray}
\begin{eqnarray}
	L_2^0(\sigma,x)=\frac{2 \Big(\sigma[\alpha ^4 (2-3 \sigma )-3 \alpha ^2 (\sigma -1)+2]+i m(2 \alpha ^3 \sigma +2 \alpha  \sigma +\alpha)+s[(\alpha ^2+1)(\alpha^2\sigma +\sigma -1)+i\alpha  x]\Big)}{4\alpha^6(\sigma-1) \sigma +\alpha^4(8 \sigma ^2-8 \sigma -4)-4 (\sigma +1)-\alpha ^2(-4 \sigma ^2+8 \sigma +x^2+7)}\, .
\end{eqnarray}

\section{Two dimensional pseudo-spectral methods}\label{two_dimensional_pseudo_spectral_method}
In this appendix, we will introduce the two-dimensional  pseudo-spectral methods to solve ODEs or PDEs following~\cite{doi:10.1137/1.9780898719598,Miguel:2023rzp}. Furthermore, PDEs with $n$ (spatial) variables are rather direct. The so-called pseudo-spectral method is to sample the solution of the differential equation on some grid. Then, one can map the differential operators into matrix operators, simplifying the original continuous problem into a simpler finite-dimensional version.

We start from an ODE associated with variable $x$ and undetermined function $f(x)$ written as 
\begin{eqnarray}\label{ODE}
	\sum_{n}L^{(n)}(x)\frac{\mathrm{d}^n}{\mathrm{d}x^n}f(x)=0\, ,
\end{eqnarray}
in which $n$ denotes the order of the derivative. To derive the discrete counterpart of Eq.(\ref{ODE}), our initial step involves selecting a set of grid points denoted by $\{x_i\}_{i=0}^{N}$. Then, the undetermined function $f(x)$ in the chosen grid is defined by $f_i=f(x_i)$. Proceed to the next step, differentiation will have a matrix representation, that is, the so-called differential matrix $\mathbf{D}$. In order to obtain the expressions of the differential matrices for different grids, one can refer to~\cite{Sheikh:2022cud,boyd2001chebyshev}. Therefore, the first derivative of $f(x)$ in this grid is
\begin{eqnarray}
	f^{\prime}(x_i)=\sum_{j}\mathbf{D}_{ij}f_j\, .
\end{eqnarray}
The $n$-th derivative of $f(x)$ is then obtained by
\begin{eqnarray}
	f^{(n)}(x_i)=\sum_j(\mathbf{D}^n)_{ij}f_j\, .
\end{eqnarray}
After the coefficient functions $L^{(n)}(x)$ is expressed as $\mathbf{L}^{(n)}_{ij}=L^{(n)}(x_i)\delta_{ij}$, one gets the discrete representation of Eq.(\ref{ODE})
\begin{eqnarray}
	\sum_{njk}\mathbf{L}^{(n)}_{ij}(\mathbf{D}^n)_{jk}f_k=0\, .
\end{eqnarray}

\begin{figure}[htbp]
	\centering
	\includegraphics[width=0.6\textwidth]{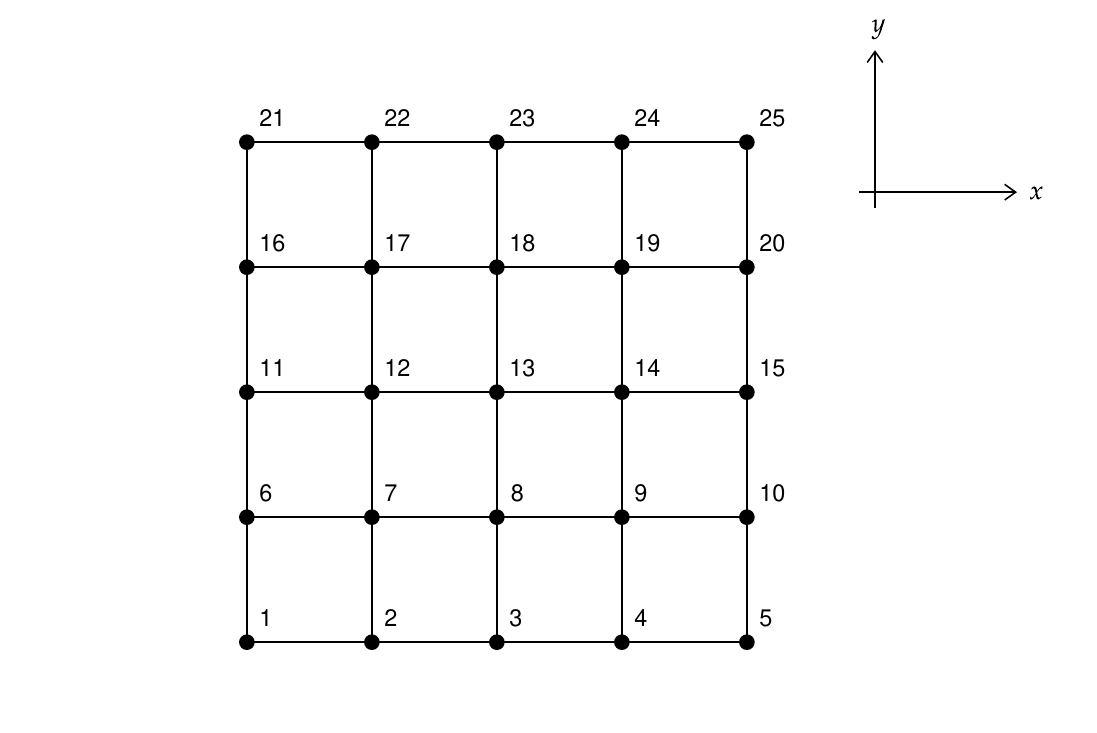}
	\caption{A $5\times 5$ tensor product grid. This is just a schematic diagram. In fact, for the Chebyshev grid, the grid is nonuniform.}
	\label{tensor_product_grid}
\end{figure}

Now, we have a PDE associated with two variables $x$, $y$ and undetermined function $f(x,y)$ written as
\begin{eqnarray}
	\sum_{mn}L^{(m,n)}(x,y)\frac{\partial^m}{\partial x^m}\frac{\partial^n}{\partial y^n}f(x,y)=0\, .
\end{eqnarray} 
In order to deal with such a PDE problem like a matrix problem, we should naturally set up a grid in each direction independently, called a tensor product grid, where each grids are denoted by $\{x_i\}_{i=0}^{N_x}$ and $\{y_{j}\}_{j=0}^{N_y}$, respectively. The function $f(x,y)$ is sampled on the grid $f_{ij}=f(x_i,y_j)$, and we use the ``lexicographic'' ordering to transform $f_{ij}$ into a vector, where such vector has dimension $(N_x+1)\times(N_y+1)$. For example, we show a $5\times 5$ tensor product grid in Fig.\ref{tensor_product_grid}. Then, the discrete versions of the following partial derivative operators become:
\begin{eqnarray}\label{partial_derivative_1}
	\frac{\partial}{\partial x}\to\mathbf{I}\otimes \mathbf{D}_x\, ,\quad\frac{\partial}{\partial y} \to\mathbf{D}_y\otimes\mathbf{I}\, ,
\end{eqnarray}
and
\begin{eqnarray}\label{partial_derivative_2}
	\frac{\partial^2}{\partial x^2}\to \mathbf{I}\otimes (\mathbf{D}_x)^2\, ,\quad \frac{\partial^2}{\partial x\partial y}\to\mathbf{D}_y\otimes\mathbf{D}_x\, ,\quad\frac{\partial^2}{\partial y^2}\to(\mathbf{D}_y)^2\otimes\mathbf{I}\, ,
\end{eqnarray}
where $\otimes$ is known as the Kronecker product of matrices and $\mathbf{I}$ is the identity matrix with dimension $N_x+1$ or $N_y+1$. It should be noted that such Kronecker products are based on the ``lexicographic'' ordering chosen in Fig.\ref{tensor_product_grid}. The coefficient functions $L^{(m,n)}(x,y)$ will become a diagonal matrix whose diagonal elements are sampled on the tensor product grid of the same order as $f_{ij}$.

\section{The construction of the Gram matrix for two dimensional inner product}\label{Gram_matrix}
In this appendix, we will introduce how one gets the Gram matrix from the two dimensional inner product. The most basic thought here is that integrals can be computed via the Clenshaw-Curtis quadrature. Given a function $f(\sigma,x)$ and a tensor product grid like Fig.\ref{tensor_product_grid}, then the integral of the function $f(\sigma,x)$ on region $[0,1]\times [-1,1]$ is approximated by
\begin{eqnarray}
    \int_0^1\mathrm{d}\sigma\int_{-1}^{1}\mathrm{d}xf(\sigma,x)\approx\sum_{i=0}^{N_{\sigma}}\sum_{j=0}^{N_x}{}^{[\sigma]}w_i{}^{[x]}w_jf(\sigma_i,x_j)\, ,
\end{eqnarray}
where ${}^{[\sigma]}w_i$ and ${}^{[x]}w_j$ are called Clenshaw-Curtis weights. If one chooses the tensor product grid coming from two Chebyshev-Gauss grids $\{\sigma_i\}_{i=0}^{N_\sigma}$ and $\{x_j\}_{j=0}^{N_x}$ with
\begin{eqnarray}
    \sigma_i=\frac{1}{2}\Big[1+\cos\Big(\frac{\pi(i+1/2)}{N_\sigma+1}\Big)\Big]\, ,\quad i=0,\cdots, N_\sigma\, ,
\end{eqnarray}
\begin{eqnarray}
    \quad x_j=\cos\Big(\frac{\pi(j+1/2)}{N_x+1}\Big)\, ,\quad j=0,\cdots, N_x\, ,
\end{eqnarray}
then two functions ${}^{[\sigma]}w_i$ and ${}^{[x]}w_j$ have following forms~\cite{Sheikh:2022cud},
\begin{eqnarray}
    {}^{[\sigma]}w_i=\frac{1}{N_\sigma+1}\Big[1-2\sum_{k=1}^{\lfloor\frac{N_\sigma}{2}\rfloor}\frac{T_{2k}(2\sigma_i-1)}{4k^2-1}\Big]\, ,\quad \text{and} \quad {}^{[x]}w_j=\frac{2}{N_x+1}\Big[1-2\sum_{k=1}^{\lfloor\frac{N_x}{2}\rfloor}\frac{T_{2k}(x_j)}{4k^2-1}\Big]\, ,
\end{eqnarray}
where $\lfloor a\rfloor$ is the floor function, i.e., the largest integer that is less than or equal to $a$, and $T_k$ is the Chebyshev polynomial of order $k$.

Now, we are going to construct the associated Gram matrix for the norms $\lVert\cdot\rVert_E$ and $\lVert\cdot\rVert_{L^2}$. For simplicity, the number of $\sigma$-grid is equal to the number of $x$-grid, i.e., $N_\sigma=N_x=N$. Using the tensor product grid, functions $V_m$ and $W_m$ are discretized and flatten into $\mathbf{V}_m$ and $\mathbf{W}_m$, in which the arrangement order is given by Fig.\ref{tensor_product_grid}. The dimensions of $\mathbf{V}_m$ and $\mathbf{W}_m$ are both $(N+1)\times(N+1)$. For the $L^2$-norm, the inner product between $u_{m1}$ and $u_{m2}$ reads
\begin{eqnarray}
    \langle \mathbf{u}_{m1},\mathbf{u}_{m2}\rangle_{L^2}=\begin{bmatrix}
        \mathbf{V}_{m1}^{\star} \mathbf{W}_{m1}^{\star}
    \end{bmatrix}
    \cdot\mathbf{G}_{L^2}\cdot
    \begin{bmatrix}
        \mathbf{V}_{m2}\\
        \mathbf{W}_{m2}
    \end{bmatrix}=
    \begin{bmatrix}
        \mathbf{V}_{m1}^{\star} \mathbf{W}_{m1}^{\star}
    \end{bmatrix}
    \cdot
    \begin{bmatrix}
        \mathbf{G}_1  & \mathbf{0}\\
        \mathbf{0}  & \mathbf{G}_2
    \end{bmatrix}
    \cdot
    \begin{bmatrix}
        \mathbf{V}_{m2}\\
        \mathbf{W}_{m2}
    \end{bmatrix}\, .
\end{eqnarray}
Here, the symbol $\star$ stands for the Hermite conjugate of vector, $\mathbf{G}_1$ and $\mathbf{G}_2$ are the matrices of order $(N+1)^2\times(N+1)^2$, and their structures are as follows
\begin{eqnarray}
    \mathbf{G}_1=\mathbf{G}_2=\mathbb{W}\, ,\quad \mathbb{W}\equiv\text{diag}({}^{[x]}w_j)\otimes\text{diag}({}^{[\sigma]}w_i)\, ,
\end{eqnarray}
where $\mathbf{G}_1$ and $\mathbf{G}_2$ are both diagonal matrices. It is worth mentioning that, considering the arrangement order in Fig.\ref{tensor_product_grid}, we should pay attention to the position of the matrix $\text{diag}({}^{[x]}w_j)$ and $\text{diag}({}^{[\sigma]}w_i)$. 

For the energy norm $\lVert\cdot\rVert_E$ and the convenience of notion, the inner product (\ref{energy_norm_inner_product}) is rewritten as
\begin{eqnarray}
    \langle u_{m1},u_{m2}\rangle_E=\int_0^1\mathrm{d}\sigma\int_{-1}^1\mathrm{d}x\Big[C_1\bar{V}_{m1}V_{m2}+C_2\bar{W}_{m1}W_{m2}+C_3(\partial_\sigma\bar{V}_{m1})(\partial_\sigma V_{m2})+C_4(\partial_x\bar{V}_{m1})(\partial_x V_{m2})\Big]\, ,
\end{eqnarray}
where the coefficient functions of each terms have been written as  $C_1$, $C_2$, $C_3$ and $C_4$, respectively. Based on similar processes, we obtain 
\begin{eqnarray}
    \langle \mathbf{u}_{m1},\mathbf{u}_{m2}\rangle_E=\begin{bmatrix}
        \mathbf{V}_{m1}^{\star} \mathbf{W}_{m1}^{\star}
    \end{bmatrix}
    \cdot\mathbf{G}_{E}\cdot
    \begin{bmatrix}
        \mathbf{V}_{m2}\\
        \mathbf{W}_{m2}
    \end{bmatrix}=
    \begin{bmatrix}
        \mathbf{V}_{m1}^{\star} \mathbf{W}_{m1}^{\star}
    \end{bmatrix}
    \cdot
    \begin{bmatrix}
        \mathbf{G}_1  & \mathbf{0}\\
        \mathbf{0}  & \mathbf{G}_2
    \end{bmatrix}
    \cdot
    \begin{bmatrix}
        \mathbf{V}_{m2}\\
        \mathbf{W}_{m2}
    \end{bmatrix}\, ,
\end{eqnarray}
where two matrices $\mathbf{G}_1$ and $\mathbf{G}_2$ are given by
\begin{eqnarray}\label{G1_E}
    \mathbf{G}_1=\mathbf{C}_1\cdot \mathbb{W}+(\mathbf{I}\otimes \mathbf{D}_\sigma)^T\cdot\mathbf{C}_3\cdot\mathbb{W}\cdot (\mathbf{I}\otimes \mathbf{D}_\sigma)+(\mathbf{D}_x\otimes\mathbf{I})^T\cdot\mathbf{C}_4\cdot\mathbb{W}\cdot (\mathbf{D}_x\otimes\mathbf{I})\, ,
\end{eqnarray}
\begin{eqnarray}\label{G2_E}
    \mathbf{G}_2=\mathbf{C}_2\cdot \mathbb{W}\, ,
\end{eqnarray}
respectively. Eq.(\ref{G1_E}) and Eq.(\ref{G2_E}) come from Eq.(\ref{partial_derivative_1}) and Eq.(\ref{partial_derivative_2}). The symbol $T$ represents the transpose of the matrix. In Eq.(\ref{G1_E}) and Eq.(\ref{G2_E}), $\mathbf{C}_1$, $\mathbf{C}_2$, $\mathbf{C}_3$ and $\mathbf{C}_4$ are the diagonal matrices of order $(N+1)^2\times(N+1)^2$ and their diagonal elements are the values of the function $C_i$, $i=1,2,3,4$ on the tensor product grid, respectively.

\section{The pseudospectra for other $\alpha$'s}\label{other_alpha}
\begin{figure}[htbp]
	\centering
    \includegraphics[width=0.4\textwidth]{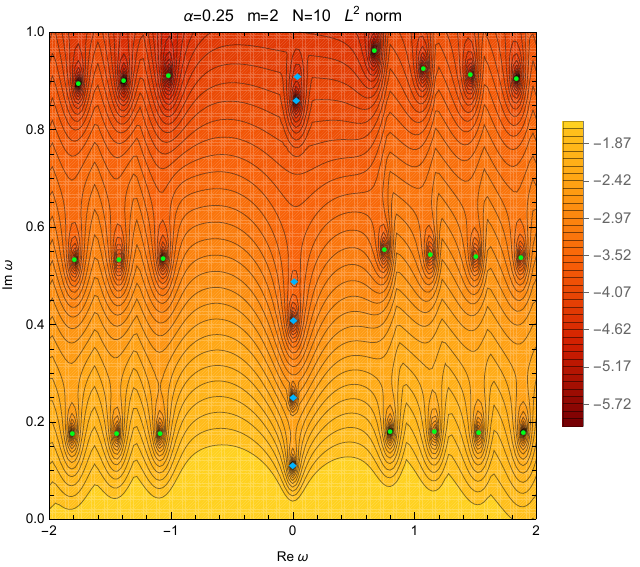}
    \hspace{0.4cm}
    \includegraphics[width=0.4\textwidth]{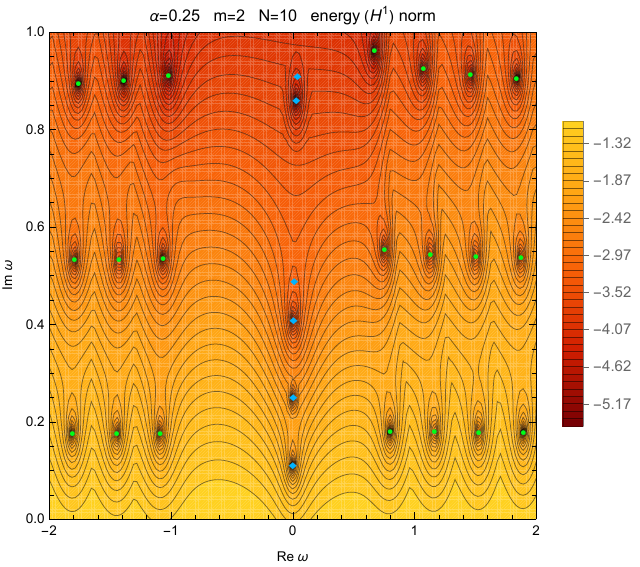}
    \includegraphics[width=0.4\textwidth]{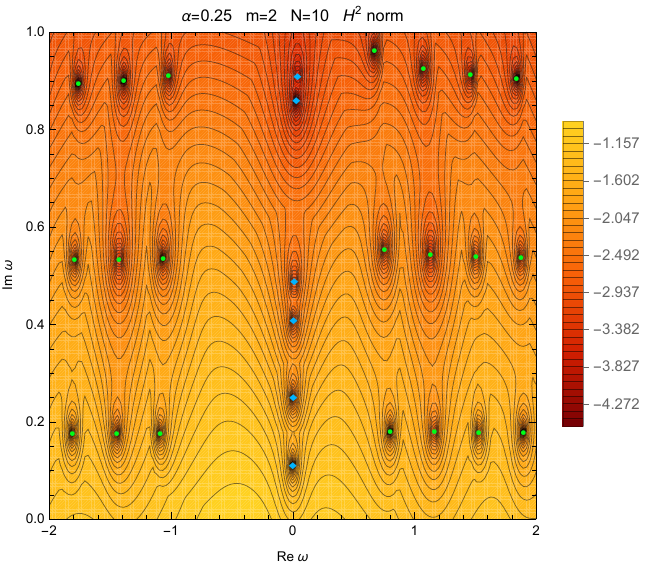}
    \hspace{0.4cm}
    \includegraphics[width=0.4\textwidth]{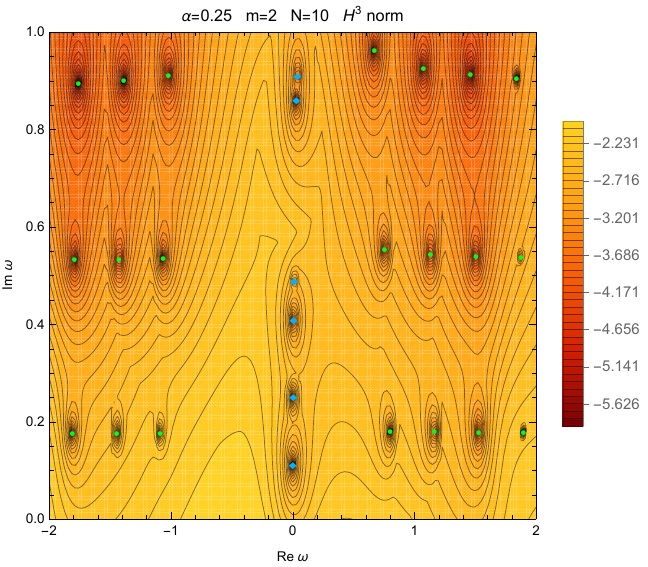}
    \includegraphics[width=0.4\textwidth]{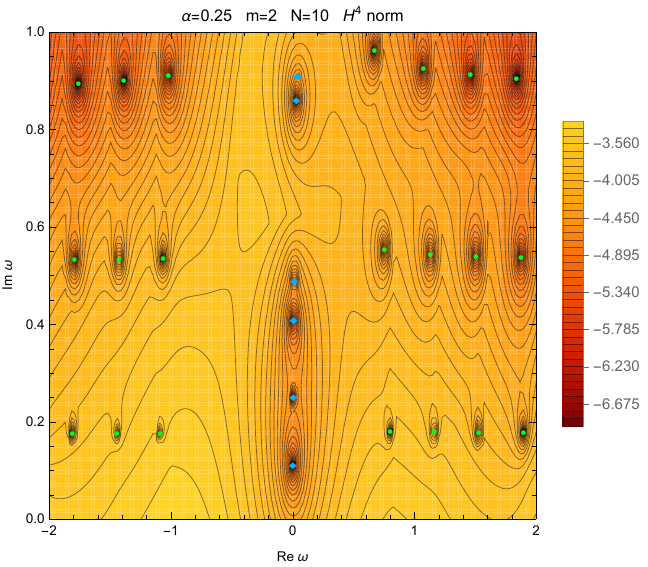}
    \hspace{0.4cm}
    \includegraphics[width=0.4\textwidth]{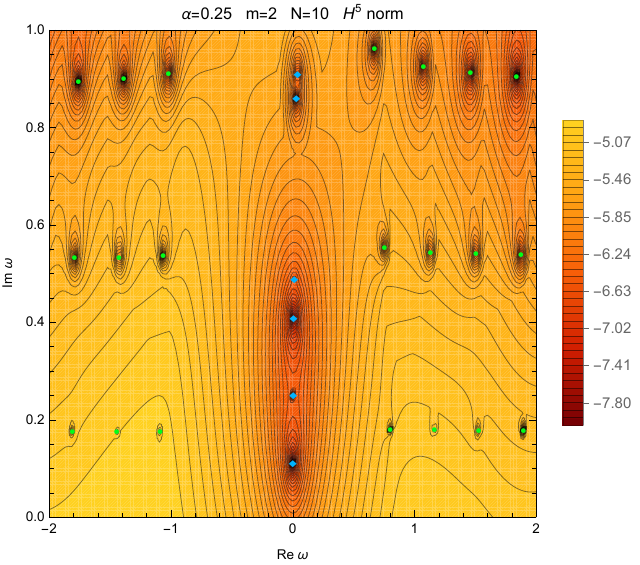}
	\caption{The pseudospectra $\log_{10}[\sigma_{\epsilon}(L)]$ for $L^2$-norm, energy ($H^1$) norm, $H^2$-norm, $H^3$-norm, $H^4$-norm and $H^5$-norm of Kerr black hole with $\alpha=0.25$ and $m=2$. Here, the resolution for the operator $L$ is given by $N=10$. For all six panels, the scopes of drawings are all limited to $\text{Re}\omega_{\text{max}}=2$, $\text{Re}\omega_{\text{min}}=-2$, $\text{Im}\omega_{\text{min}}=0$ and $\text{Im}\omega_{\text{max}}=1$. The resolutions of pseudospectra figures are all set as $\Delta\text{Re}\omega=4/150$ and $\Delta\text{Im}\omega=1/150$. In addition, the number of contour lines is $40$ for these six panels.}
    \label{pseudospectra_alpha_1_4_m_2}
\end{figure}

In this appendix, we show the pseudospectra for the case $\alpha=0.25$ with several norms defined in Sec.\ref{norm} in Fig.\ref{pseudospectra_alpha_1_4_m_2}. 

\bibliography{reference}{}

\begin{thebibliography}{85}%
\makeatletter
\providecommand \@ifxundefined [1]{%
 \@ifx{#1\undefined}
}%
\providecommand \@ifnum [1]{%
 \ifnum #1\expandafter \@firstoftwo
 \else \expandafter \@secondoftwo
 \fi
}%
\providecommand \@ifx [1]{%
 \ifx #1\expandafter \@firstoftwo
 \else \expandafter \@secondoftwo
 \fi
}%
\providecommand \natexlab [1]{#1}%
\providecommand \enquote  [1]{``#1''}%
\providecommand \bibnamefont  [1]{#1}%
\providecommand \bibfnamefont [1]{#1}%
\providecommand \citenamefont [1]{#1}%
\providecommand \href@noop [0]{\@secondoftwo}%
\providecommand \href [0]{\begingroup \@sanitize@url \@href}%
\providecommand \@href[1]{\@@startlink{#1}\@@href}%
\providecommand \@@href[1]{\endgroup#1\@@endlink}%
\providecommand \@sanitize@url [0]{\catcode `\\12\catcode `\$12\catcode
  `\&12\catcode `\#12\catcode `\^12\catcode `\_12\catcode `\%12\relax}%
\providecommand \@@startlink[1]{}%
\providecommand \@@endlink[0]{}%
\providecommand \url  [0]{\begingroup\@sanitize@url \@url }%
\providecommand \@url [1]{\endgroup\@href {#1}{\urlprefix }}%
\providecommand \urlprefix  [0]{URL }%
\providecommand \Eprint [0]{\href }%
\providecommand \doibase [0]{http://dx.doi.org/}%
\providecommand \selectlanguage [0]{\@gobble}%
\providecommand \bibinfo  [0]{\@secondoftwo}%
\providecommand \bibfield  [0]{\@secondoftwo}%
\providecommand \translation [1]{[#1]}%
\providecommand \BibitemOpen [0]{}%
\providecommand \bibitemStop [0]{}%
\providecommand \bibitemNoStop [0]{.\EOS\space}%
\providecommand \EOS [0]{\spacefactor3000\relax}%
\providecommand \BibitemShut  [1]{\csname bibitem#1\endcsname}%
\let\auto@bib@innerbib\@empty
%</preamble>
\bibitem [{\citenamefont {Teukolsky}(1972)}]{Teukolsky:1972my}%
  \BibitemOpen
  \bibfield  {author} {\bibinfo {author} {\bibfnamefont {S.~A.}\ \bibnamefont
  {Teukolsky}},\ }\href {\doibase 10.1103/PhysRevLett.29.1114} {\bibfield
  {journal} {\bibinfo  {journal} {Phys. Rev. Lett.}\ }\textbf {\bibinfo
  {volume} {29}},\ \bibinfo {pages} {1114} (\bibinfo {year}
  {1972})}\BibitemShut {NoStop}%
\bibitem [{\citenamefont {Teukolsky}(1973)}]{Teukolsky:1973ha}%
  \BibitemOpen
  \bibfield  {author} {\bibinfo {author} {\bibfnamefont {S.~A.}\ \bibnamefont
  {Teukolsky}},\ }\href {\doibase 10.1086/152444} {\bibfield  {journal}
  {\bibinfo  {journal} {Astrophys. J.}\ }\textbf {\bibinfo {volume} {185}},\
  \bibinfo {pages} {635} (\bibinfo {year} {1973})}\BibitemShut {NoStop}%
\bibitem [{\citenamefont {Baibhav}\ \emph {et~al.}(2023)\citenamefont
  {Baibhav}, \citenamefont {Cheung}, \citenamefont {Berti}, \citenamefont
  {Cardoso}, \citenamefont {Carullo}, \citenamefont {Cotesta}, \citenamefont
  {Del~Pozzo},\ and\ \citenamefont {Duque}}]{Baibhav:2023clw}%
  \BibitemOpen
  \bibfield  {author} {\bibinfo {author} {\bibfnamefont {V.}~\bibnamefont
  {Baibhav}}, \bibinfo {author} {\bibfnamefont {M.~H.-Y.}\ \bibnamefont
  {Cheung}}, \bibinfo {author} {\bibfnamefont {E.}~\bibnamefont {Berti}},
  \bibinfo {author} {\bibfnamefont {V.}~\bibnamefont {Cardoso}}, \bibinfo
  {author} {\bibfnamefont {G.}~\bibnamefont {Carullo}}, \bibinfo {author}
  {\bibfnamefont {R.}~\bibnamefont {Cotesta}}, \bibinfo {author} {\bibfnamefont
  {W.}~\bibnamefont {Del~Pozzo}}, \ and\ \bibinfo {author} {\bibfnamefont
  {F.}~\bibnamefont {Duque}},\ }\href {\doibase 10.1103/PhysRevD.108.104020}
  {\bibfield  {journal} {\bibinfo  {journal} {Phys. Rev. D}\ }\textbf {\bibinfo
  {volume} {108}},\ \bibinfo {pages} {104020} (\bibinfo {year} {2023})},\
  \Eprint {http://arxiv.org/abs/2302.03050} {arXiv:2302.03050 [gr-qc]}
  \BibitemShut {NoStop}%
\bibitem [{\citenamefont {Giesler}\ \emph {et~al.}(2019)\citenamefont
  {Giesler}, \citenamefont {Isi}, \citenamefont {Scheel},\ and\ \citenamefont
  {Teukolsky}}]{Giesler:2019uxc}%
  \BibitemOpen
  \bibfield  {author} {\bibinfo {author} {\bibfnamefont {M.}~\bibnamefont
  {Giesler}}, \bibinfo {author} {\bibfnamefont {M.}~\bibnamefont {Isi}},
  \bibinfo {author} {\bibfnamefont {M.~A.}\ \bibnamefont {Scheel}}, \ and\
  \bibinfo {author} {\bibfnamefont {S.}~\bibnamefont {Teukolsky}},\ }\href
  {\doibase 10.1103/PhysRevX.9.041060} {\bibfield  {journal} {\bibinfo
  {journal} {Phys. Rev. X}\ }\textbf {\bibinfo {volume} {9}},\ \bibinfo {pages}
  {041060} (\bibinfo {year} {2019})},\ \Eprint
  {http://arxiv.org/abs/1903.08284} {arXiv:1903.08284 [gr-qc]} \BibitemShut
  {NoStop}%
\bibitem [{\citenamefont {Barausse}\ \emph {et~al.}(2014)\citenamefont
  {Barausse}, \citenamefont {Cardoso},\ and\ \citenamefont
  {Pani}}]{Barausse:2014tra}%
  \BibitemOpen
  \bibfield  {author} {\bibinfo {author} {\bibfnamefont {E.}~\bibnamefont
  {Barausse}}, \bibinfo {author} {\bibfnamefont {V.}~\bibnamefont {Cardoso}}, \
  and\ \bibinfo {author} {\bibfnamefont {P.}~\bibnamefont {Pani}},\ }\href
  {\doibase 10.1103/PhysRevD.89.104059} {\bibfield  {journal} {\bibinfo
  {journal} {Phys. Rev. D}\ }\textbf {\bibinfo {volume} {89}},\ \bibinfo
  {pages} {104059} (\bibinfo {year} {2014})},\ \Eprint
  {http://arxiv.org/abs/1404.7149} {arXiv:1404.7149 [gr-qc]} \BibitemShut
  {NoStop}%
\bibitem [{\citenamefont {Nollert}(1996)}]{Nollert:1996rf}%
  \BibitemOpen
  \bibfield  {author} {\bibinfo {author} {\bibfnamefont {H.-P.}\ \bibnamefont
  {Nollert}},\ }\href {\doibase 10.1103/PhysRevD.53.4397} {\bibfield  {journal}
  {\bibinfo  {journal} {Phys. Rev. D}\ }\textbf {\bibinfo {volume} {53}},\
  \bibinfo {pages} {4397} (\bibinfo {year} {1996})},\ \Eprint
  {http://arxiv.org/abs/gr-qc/9602032} {arXiv:gr-qc/9602032} \BibitemShut
  {NoStop}%
\bibitem [{\citenamefont {Nollert}\ and\ \citenamefont
  {Price}(1999)}]{Nollert:1998ys}%
  \BibitemOpen
  \bibfield  {author} {\bibinfo {author} {\bibfnamefont {H.-P.}\ \bibnamefont
  {Nollert}}\ and\ \bibinfo {author} {\bibfnamefont {R.~H.}\ \bibnamefont
  {Price}},\ }\href {\doibase 10.1063/1.532698} {\bibfield  {journal} {\bibinfo
   {journal} {J. Math. Phys.}\ }\textbf {\bibinfo {volume} {40}},\ \bibinfo
  {pages} {980} (\bibinfo {year} {1999})},\ \Eprint
  {http://arxiv.org/abs/gr-qc/9810074} {arXiv:gr-qc/9810074} \BibitemShut
  {NoStop}%
\bibitem [{\citenamefont {Daghigh}\ \emph {et~al.}(2020)\citenamefont
  {Daghigh}, \citenamefont {Green},\ and\ \citenamefont
  {Morey}}]{Daghigh:2020jyk}%
  \BibitemOpen
  \bibfield  {author} {\bibinfo {author} {\bibfnamefont {R.~G.}\ \bibnamefont
  {Daghigh}}, \bibinfo {author} {\bibfnamefont {M.~D.}\ \bibnamefont {Green}},
  \ and\ \bibinfo {author} {\bibfnamefont {J.~C.}\ \bibnamefont {Morey}},\
  }\href {\doibase 10.1103/PhysRevD.101.104009} {\bibfield  {journal} {\bibinfo
   {journal} {Phys. Rev. D}\ }\textbf {\bibinfo {volume} {101}},\ \bibinfo
  {pages} {104009} (\bibinfo {year} {2020})},\ \Eprint
  {http://arxiv.org/abs/2002.07251} {arXiv:2002.07251 [gr-qc]} \BibitemShut
  {NoStop}%
\bibitem [{\citenamefont {Qian}\ \emph {et~al.}(2021)\citenamefont {Qian},
  \citenamefont {Lin}, \citenamefont {Shao}, \citenamefont {Wang},\ and\
  \citenamefont {Yue}}]{Qian:2020cnz}%
  \BibitemOpen
  \bibfield  {author} {\bibinfo {author} {\bibfnamefont {W.-L.}\ \bibnamefont
  {Qian}}, \bibinfo {author} {\bibfnamefont {K.}~\bibnamefont {Lin}}, \bibinfo
  {author} {\bibfnamefont {C.-Y.}\ \bibnamefont {Shao}}, \bibinfo {author}
  {\bibfnamefont {B.}~\bibnamefont {Wang}}, \ and\ \bibinfo {author}
  {\bibfnamefont {R.-H.}\ \bibnamefont {Yue}},\ }\href {\doibase
  10.1103/PhysRevD.103.024019} {\bibfield  {journal} {\bibinfo  {journal}
  {Phys. Rev. D}\ }\textbf {\bibinfo {volume} {103}},\ \bibinfo {pages}
  {024019} (\bibinfo {year} {2021})},\ \Eprint
  {http://arxiv.org/abs/2009.11627} {arXiv:2009.11627 [gr-qc]} \BibitemShut
  {NoStop}%
\bibitem [{\citenamefont {Berti}\ \emph {et~al.}(2022)\citenamefont {Berti},
  \citenamefont {Cardoso}, \citenamefont {Cheung}, \citenamefont {Di~Filippo},
  \citenamefont {Duque}, \citenamefont {Martens},\ and\ \citenamefont
  {Mukohyama}}]{Berti:2022xfj}%
  \BibitemOpen
  \bibfield  {author} {\bibinfo {author} {\bibfnamefont {E.}~\bibnamefont
  {Berti}}, \bibinfo {author} {\bibfnamefont {V.}~\bibnamefont {Cardoso}},
  \bibinfo {author} {\bibfnamefont {M.~H.-Y.}\ \bibnamefont {Cheung}}, \bibinfo
  {author} {\bibfnamefont {F.}~\bibnamefont {Di~Filippo}}, \bibinfo {author}
  {\bibfnamefont {F.}~\bibnamefont {Duque}}, \bibinfo {author} {\bibfnamefont
  {P.}~\bibnamefont {Martens}}, \ and\ \bibinfo {author} {\bibfnamefont
  {S.}~\bibnamefont {Mukohyama}},\ }\href {\doibase
  10.1103/PhysRevD.106.084011} {\bibfield  {journal} {\bibinfo  {journal}
  {Phys. Rev. D}\ }\textbf {\bibinfo {volume} {106}},\ \bibinfo {pages}
  {084011} (\bibinfo {year} {2022})},\ \Eprint
  {http://arxiv.org/abs/2205.08547} {arXiv:2205.08547 [gr-qc]} \BibitemShut
  {NoStop}%
\bibitem [{\citenamefont {Cheung}\ \emph {et~al.}(2022)\citenamefont {Cheung},
  \citenamefont {Destounis}, \citenamefont {Macedo}, \citenamefont {Berti},\
  and\ \citenamefont {Cardoso}}]{Cheung:2021bol}%
  \BibitemOpen
  \bibfield  {author} {\bibinfo {author} {\bibfnamefont {M.~H.-Y.}\
  \bibnamefont {Cheung}}, \bibinfo {author} {\bibfnamefont {K.}~\bibnamefont
  {Destounis}}, \bibinfo {author} {\bibfnamefont {R.~P.}\ \bibnamefont
  {Macedo}}, \bibinfo {author} {\bibfnamefont {E.}~\bibnamefont {Berti}}, \
  and\ \bibinfo {author} {\bibfnamefont {V.}~\bibnamefont {Cardoso}},\ }\href
  {\doibase 10.1103/PhysRevLett.128.111103} {\bibfield  {journal} {\bibinfo
  {journal} {Phys. Rev. Lett.}\ }\textbf {\bibinfo {volume} {128}},\ \bibinfo
  {pages} {111103} (\bibinfo {year} {2022})},\ \Eprint
  {http://arxiv.org/abs/2111.05415} {arXiv:2111.05415 [gr-qc]} \BibitemShut
  {NoStop}%
\bibitem [{\citenamefont {Li}\ \emph {et~al.}(2024)\citenamefont {Li},
  \citenamefont {Qian},\ and\ \citenamefont {Daghigh}}]{Li:2024npg}%
  \BibitemOpen
  \bibfield  {author} {\bibinfo {author} {\bibfnamefont {G.-R.}\ \bibnamefont
  {Li}}, \bibinfo {author} {\bibfnamefont {W.-L.}\ \bibnamefont {Qian}}, \ and\
  \bibinfo {author} {\bibfnamefont {R.~G.}\ \bibnamefont {Daghigh}},\ }\href
  {\doibase 10.1103/PhysRevD.110.064076} {\bibfield  {journal} {\bibinfo
  {journal} {Phys. Rev. D}\ }\textbf {\bibinfo {volume} {110}},\ \bibinfo
  {pages} {064076} (\bibinfo {year} {2024})},\ \Eprint
  {http://arxiv.org/abs/2406.10782} {arXiv:2406.10782 [gr-qc]} \BibitemShut
  {NoStop}%
\bibitem [{\citenamefont {Yang}\ \emph {et~al.}(2024)\citenamefont {Yang},
  \citenamefont {Mai}, \citenamefont {Yang}, \citenamefont {Shao},\ and\
  \citenamefont {Berti}}]{Yang:2024vor}%
  \BibitemOpen
  \bibfield  {author} {\bibinfo {author} {\bibfnamefont {Y.}~\bibnamefont
  {Yang}}, \bibinfo {author} {\bibfnamefont {Z.-F.}\ \bibnamefont {Mai}},
  \bibinfo {author} {\bibfnamefont {R.-Q.}\ \bibnamefont {Yang}}, \bibinfo
  {author} {\bibfnamefont {L.}~\bibnamefont {Shao}}, \ and\ \bibinfo {author}
  {\bibfnamefont {E.}~\bibnamefont {Berti}},\ }\href {\doibase
  10.1103/PhysRevD.110.084018} {\bibfield  {journal} {\bibinfo  {journal}
  {Phys. Rev. D}\ }\textbf {\bibinfo {volume} {110}},\ \bibinfo {pages}
  {084018} (\bibinfo {year} {2024})},\ \Eprint
  {http://arxiv.org/abs/2407.20131} {arXiv:2407.20131 [gr-qc]} \BibitemShut
  {NoStop}%
\bibitem [{\citenamefont {Courty}\ \emph {et~al.}(2023)\citenamefont {Courty},
  \citenamefont {Destounis},\ and\ \citenamefont {Pani}}]{Courty:2023rxk}%
  \BibitemOpen
  \bibfield  {author} {\bibinfo {author} {\bibfnamefont {A.}~\bibnamefont
  {Courty}}, \bibinfo {author} {\bibfnamefont {K.}~\bibnamefont {Destounis}}, \
  and\ \bibinfo {author} {\bibfnamefont {P.}~\bibnamefont {Pani}},\ }\href
  {\doibase 10.1103/PhysRevD.108.104027} {\bibfield  {journal} {\bibinfo
  {journal} {Phys. Rev. D}\ }\textbf {\bibinfo {volume} {108}},\ \bibinfo
  {pages} {104027} (\bibinfo {year} {2023})},\ \Eprint
  {http://arxiv.org/abs/2307.11155} {arXiv:2307.11155 [gr-qc]} \BibitemShut
  {NoStop}%
\bibitem [{\citenamefont {Cardoso}\ \emph {et~al.}(2024)\citenamefont
  {Cardoso}, \citenamefont {Kastha},\ and\ \citenamefont
  {Panosso~Macedo}}]{Cardoso:2024mrw}%
  \BibitemOpen
  \bibfield  {author} {\bibinfo {author} {\bibfnamefont {V.}~\bibnamefont
  {Cardoso}}, \bibinfo {author} {\bibfnamefont {S.}~\bibnamefont {Kastha}}, \
  and\ \bibinfo {author} {\bibfnamefont {R.}~\bibnamefont {Panosso~Macedo}},\
  }\href {\doibase 10.1103/PhysRevD.110.024016} {\bibfield  {journal} {\bibinfo
   {journal} {Phys. Rev. D}\ }\textbf {\bibinfo {volume} {110}},\ \bibinfo
  {pages} {024016} (\bibinfo {year} {2024})},\ \Eprint
  {http://arxiv.org/abs/2404.01374} {arXiv:2404.01374 [gr-qc]} \BibitemShut
  {NoStop}%
\bibitem [{\citenamefont {Ianniccari}\ \emph {et~al.}(2024)\citenamefont
  {Ianniccari}, \citenamefont {Iovino}, \citenamefont {Kehagias}, \citenamefont
  {Pani}, \citenamefont {Perna}, \citenamefont {Perrone},\ and\ \citenamefont
  {Riotto}}]{Ianniccari:2024ysv}%
  \BibitemOpen
  \bibfield  {author} {\bibinfo {author} {\bibfnamefont {A.}~\bibnamefont
  {Ianniccari}}, \bibinfo {author} {\bibfnamefont {A.~J.}\ \bibnamefont
  {Iovino}}, \bibinfo {author} {\bibfnamefont {A.}~\bibnamefont {Kehagias}},
  \bibinfo {author} {\bibfnamefont {P.}~\bibnamefont {Pani}}, \bibinfo {author}
  {\bibfnamefont {G.}~\bibnamefont {Perna}}, \bibinfo {author} {\bibfnamefont
  {D.}~\bibnamefont {Perrone}}, \ and\ \bibinfo {author} {\bibfnamefont
  {A.}~\bibnamefont {Riotto}},\ }\href {\doibase
  10.1103/PhysRevLett.133.211401} {\bibfield  {journal} {\bibinfo  {journal}
  {Phys. Rev. Lett.}\ }\textbf {\bibinfo {volume} {133}},\ \bibinfo {pages}
  {211401} (\bibinfo {year} {2024})},\ \Eprint
  {http://arxiv.org/abs/2407.20144} {arXiv:2407.20144 [gr-qc]} \BibitemShut
  {NoStop}%
\bibitem [{\citenamefont {Oshita}(2024)}]{Oshita:2023cjz}%
  \BibitemOpen
  \bibfield  {author} {\bibinfo {author} {\bibfnamefont {N.}~\bibnamefont
  {Oshita}},\ }\href {\doibase 10.1103/PhysRevD.109.104028} {\bibfield
  {journal} {\bibinfo  {journal} {Phys. Rev. D}\ }\textbf {\bibinfo {volume}
  {109}},\ \bibinfo {pages} {104028} (\bibinfo {year} {2024})},\ \Eprint
  {http://arxiv.org/abs/2309.05725} {arXiv:2309.05725 [gr-qc]} \BibitemShut
  {NoStop}%
\bibitem [{\citenamefont {Rosato}\ \emph {et~al.}(2024)\citenamefont {Rosato},
  \citenamefont {Destounis},\ and\ \citenamefont {Pani}}]{Rosato:2024arw}%
  \BibitemOpen
  \bibfield  {author} {\bibinfo {author} {\bibfnamefont {R.~F.}\ \bibnamefont
  {Rosato}}, \bibinfo {author} {\bibfnamefont {K.}~\bibnamefont {Destounis}}, \
  and\ \bibinfo {author} {\bibfnamefont {P.}~\bibnamefont {Pani}},\ }\href@noop
  {} {\  (\bibinfo {year} {2024})},\ \Eprint {http://arxiv.org/abs/2406.01692}
  {arXiv:2406.01692 [gr-qc]} \BibitemShut {NoStop}%
\bibitem [{\citenamefont {Oshita}\ \emph {et~al.}(2024)\citenamefont {Oshita},
  \citenamefont {Takahashi},\ and\ \citenamefont {Mukohyama}}]{Oshita:2024fzf}%
  \BibitemOpen
  \bibfield  {author} {\bibinfo {author} {\bibfnamefont {N.}~\bibnamefont
  {Oshita}}, \bibinfo {author} {\bibfnamefont {K.}~\bibnamefont {Takahashi}}, \
  and\ \bibinfo {author} {\bibfnamefont {S.}~\bibnamefont {Mukohyama}},\ }\href
  {\doibase 10.1103/PhysRevD.110.084070} {\bibfield  {journal} {\bibinfo
  {journal} {Phys. Rev. D}\ }\textbf {\bibinfo {volume} {110}},\ \bibinfo
  {pages} {084070} (\bibinfo {year} {2024})},\ \Eprint
  {http://arxiv.org/abs/2406.04525} {arXiv:2406.04525 [gr-qc]} \BibitemShut
  {NoStop}%
\bibitem [{\citenamefont {Wu}\ \emph {et~al.}(2024)\citenamefont {Wu},
  \citenamefont {Cai},\ and\ \citenamefont {Xie}}]{Wu:2024ldo}%
  \BibitemOpen
  \bibfield  {author} {\bibinfo {author} {\bibfnamefont {L.-B.}\ \bibnamefont
  {Wu}}, \bibinfo {author} {\bibfnamefont {R.-G.}\ \bibnamefont {Cai}}, \ and\
  \bibinfo {author} {\bibfnamefont {L.}~\bibnamefont {Xie}},\ }\href@noop {} {\
   (\bibinfo {year} {2024})},\ \Eprint {http://arxiv.org/abs/2411.07734}
  {arXiv:2411.07734 [gr-qc]} \BibitemShut {NoStop}%
\bibitem [{\citenamefont {Torres}(2023)}]{Torres:2023nqg}%
  \BibitemOpen
  \bibfield  {author} {\bibinfo {author} {\bibfnamefont {T.}~\bibnamefont
  {Torres}},\ }\href {\doibase 10.1103/PhysRevLett.131.111401} {\bibfield
  {journal} {\bibinfo  {journal} {Phys. Rev. Lett.}\ }\textbf {\bibinfo
  {volume} {131}},\ \bibinfo {pages} {111401} (\bibinfo {year} {2023})},\
  \Eprint {http://arxiv.org/abs/2304.10252} {arXiv:2304.10252 [gr-qc]}
  \BibitemShut {NoStop}%
\bibitem [{\citenamefont {Trefethen}\ and\ \citenamefont
  {Embree}(2005)}]{trefethen2020spectra}%
  \BibitemOpen
  \bibfield  {author} {\bibinfo {author} {\bibfnamefont {L.}~\bibnamefont
  {Trefethen}}\ and\ \bibinfo {author} {\bibfnamefont {M.}~\bibnamefont
  {Embree}},\ }\href {\doibase 10.2307/j.ctvzxx9kj} {\emph {\bibinfo {title}
  {Spectra and Pseudospectra: The Behavior of Nonnormal Matrices and
  Operators}}}\ (\bibinfo  {publisher} {Princeton university press},\ \bibinfo
  {year} {2005})\BibitemShut {NoStop}%
\bibitem [{\citenamefont {Boyanov}(2024)}]{Boyanov:2024fgc}%
  \BibitemOpen
  \bibfield  {author} {\bibinfo {author} {\bibfnamefont {V.}~\bibnamefont
  {Boyanov}}\ }(\bibinfo {year} {2024})\ \Eprint
  {http://arxiv.org/abs/2410.11547} {arXiv:2410.11547 [gr-qc]} \BibitemShut
  {NoStop}%
\bibitem [{\citenamefont {Jaramillo}\ \emph {et~al.}(2021)\citenamefont
  {Jaramillo}, \citenamefont {Panosso~Macedo},\ and\ \citenamefont
  {Al~Sheikh}}]{Jaramillo:2020tuu}%
  \BibitemOpen
  \bibfield  {author} {\bibinfo {author} {\bibfnamefont {J.~L.}\ \bibnamefont
  {Jaramillo}}, \bibinfo {author} {\bibfnamefont {R.}~\bibnamefont
  {Panosso~Macedo}}, \ and\ \bibinfo {author} {\bibfnamefont {L.}~\bibnamefont
  {Al~Sheikh}},\ }\href {\doibase 10.1103/PhysRevX.11.031003} {\bibfield
  {journal} {\bibinfo  {journal} {Phys. Rev. X}\ }\textbf {\bibinfo {volume}
  {11}},\ \bibinfo {pages} {031003} (\bibinfo {year} {2021})},\ \Eprint
  {http://arxiv.org/abs/2004.06434} {arXiv:2004.06434 [gr-qc]} \BibitemShut
  {NoStop}%
\bibitem [{\citenamefont {Destounis}\ and\ \citenamefont
  {Duque}(2023)}]{Destounis:2023ruj}%
  \BibitemOpen
  \bibfield  {author} {\bibinfo {author} {\bibfnamefont {K.}~\bibnamefont
  {Destounis}}\ and\ \bibinfo {author} {\bibfnamefont {F.}~\bibnamefont
  {Duque}}\ }(\bibinfo {year} {2023})\ \Eprint
  {http://arxiv.org/abs/2308.16227} {arXiv:2308.16227 [gr-qc]} \BibitemShut
  {NoStop}%
\bibitem [{\citenamefont {Jaramillo}\ \emph {et~al.}(2022)\citenamefont
  {Jaramillo}, \citenamefont {Panosso~Macedo},\ and\ \citenamefont
  {Sheikh}}]{Jaramillo:2021tmt}%
  \BibitemOpen
  \bibfield  {author} {\bibinfo {author} {\bibfnamefont {J.~L.}\ \bibnamefont
  {Jaramillo}}, \bibinfo {author} {\bibfnamefont {R.}~\bibnamefont
  {Panosso~Macedo}}, \ and\ \bibinfo {author} {\bibfnamefont {L.~A.}\
  \bibnamefont {Sheikh}},\ }\href {\doibase 10.1103/PhysRevLett.128.211102}
  {\bibfield  {journal} {\bibinfo  {journal} {Phys. Rev. Lett.}\ }\textbf
  {\bibinfo {volume} {128}},\ \bibinfo {pages} {211102} (\bibinfo {year}
  {2022})},\ \Eprint {http://arxiv.org/abs/2105.03451} {arXiv:2105.03451
  [gr-qc]} \BibitemShut {NoStop}%
\bibitem [{\citenamefont {Trefethen}\ \emph {et~al.}(1993)\citenamefont
  {Trefethen}, \citenamefont {Trefethen}, \citenamefont {Reddy},\ and\
  \citenamefont {Driscoll}}]{science.261.5121.578}%
  \BibitemOpen
  \bibfield  {author} {\bibinfo {author} {\bibfnamefont {L.~N.}\ \bibnamefont
  {Trefethen}}, \bibinfo {author} {\bibfnamefont {A.~E.}\ \bibnamefont
  {Trefethen}}, \bibinfo {author} {\bibfnamefont {S.~C.}\ \bibnamefont
  {Reddy}}, \ and\ \bibinfo {author} {\bibfnamefont {T.~A.}\ \bibnamefont
  {Driscoll}},\ }\href {\doibase 10.1126/science.261.5121.578} {\bibfield
  {journal} {\bibinfo  {journal} {Science}\ }\textbf {\bibinfo {volume}
  {261}},\ \bibinfo {pages} {578} (\bibinfo {year} {1993})}\BibitemShut
  {NoStop}%
\bibitem [{\citenamefont {Destounis}\ \emph {et~al.}(2021)\citenamefont
  {Destounis}, \citenamefont {Macedo}, \citenamefont {Berti}, \citenamefont
  {Cardoso},\ and\ \citenamefont {Jaramillo}}]{Destounis:2021lum}%
  \BibitemOpen
  \bibfield  {author} {\bibinfo {author} {\bibfnamefont {K.}~\bibnamefont
  {Destounis}}, \bibinfo {author} {\bibfnamefont {R.~P.}\ \bibnamefont
  {Macedo}}, \bibinfo {author} {\bibfnamefont {E.}~\bibnamefont {Berti}},
  \bibinfo {author} {\bibfnamefont {V.}~\bibnamefont {Cardoso}}, \ and\
  \bibinfo {author} {\bibfnamefont {J.~L.}\ \bibnamefont {Jaramillo}},\ }\href
  {\doibase 10.1103/PhysRevD.104.084091} {\bibfield  {journal} {\bibinfo
  {journal} {Phys. Rev. D}\ }\textbf {\bibinfo {volume} {104}},\ \bibinfo
  {pages} {084091} (\bibinfo {year} {2021})},\ \Eprint
  {http://arxiv.org/abs/2107.09673} {arXiv:2107.09673 [gr-qc]} \BibitemShut
  {NoStop}%
\bibitem [{\citenamefont {Cao}\ \emph {et~al.}(2024{\natexlab{a}})\citenamefont
  {Cao}, \citenamefont {Chen}, \citenamefont {Wu}, \citenamefont {Xie},\ and\
  \citenamefont {Zhou}}]{Cao:2024oud}%
  \BibitemOpen
  \bibfield  {author} {\bibinfo {author} {\bibfnamefont {L.-M.}\ \bibnamefont
  {Cao}}, \bibinfo {author} {\bibfnamefont {J.-N.}\ \bibnamefont {Chen}},
  \bibinfo {author} {\bibfnamefont {L.-B.}\ \bibnamefont {Wu}}, \bibinfo
  {author} {\bibfnamefont {L.}~\bibnamefont {Xie}}, \ and\ \bibinfo {author}
  {\bibfnamefont {Y.-S.}\ \bibnamefont {Zhou}},\ }\href {\doibase
  10.1007/s11433-024-2435-5} {\bibfield  {journal} {\bibinfo  {journal} {Sci.
  China Phys. Mech. Astron.}\ }\textbf {\bibinfo {volume} {67}},\ \bibinfo
  {pages} {100412} (\bibinfo {year} {2024}{\natexlab{a}})},\ \Eprint
  {http://arxiv.org/abs/2401.09907} {arXiv:2401.09907 [gr-qc]} \BibitemShut
  {NoStop}%
\bibitem [{\citenamefont {Cao}\ \emph {et~al.}(2024{\natexlab{b}})\citenamefont
  {Cao}, \citenamefont {Wu},\ and\ \citenamefont {Zhou}}]{Cao:2024sot}%
  \BibitemOpen
  \bibfield  {author} {\bibinfo {author} {\bibfnamefont {L.-M.}\ \bibnamefont
  {Cao}}, \bibinfo {author} {\bibfnamefont {L.-B.}\ \bibnamefont {Wu}}, \ and\
  \bibinfo {author} {\bibfnamefont {Y.-S.}\ \bibnamefont {Zhou}},\ }\href@noop
  {} {\  (\bibinfo {year} {2024}{\natexlab{b}})},\ \Eprint
  {http://arxiv.org/abs/2412.21092} {arXiv:2412.21092 [gr-qc]} \BibitemShut
  {NoStop}%
\bibitem [{\citenamefont {Arean}\ \emph {et~al.}(2024)\citenamefont {Arean},
  \citenamefont {Garcia-Fari\~na},\ and\ \citenamefont
  {Landsteiner}}]{Arean:2024afl}%
  \BibitemOpen
  \bibfield  {author} {\bibinfo {author} {\bibfnamefont {D.}~\bibnamefont
  {Arean}}, \bibinfo {author} {\bibfnamefont {D.}~\bibnamefont
  {Garcia-Fari\~na}}, \ and\ \bibinfo {author} {\bibfnamefont {K.}~\bibnamefont
  {Landsteiner}},\ }\href {\doibase 10.3389/fphy.2024.1460268} {\bibfield
  {journal} {\bibinfo  {journal} {Front. in Phys.}\ }\textbf {\bibinfo {volume}
  {12}},\ \bibinfo {pages} {1460268} (\bibinfo {year} {2024})},\ \Eprint
  {http://arxiv.org/abs/2407.04372} {arXiv:2407.04372 [hep-th]} \BibitemShut
  {NoStop}%
\bibitem [{\citenamefont {Garcia-Fari\~na}\ \emph {et~al.}(2024)\citenamefont
  {Garcia-Fari\~na}, \citenamefont {Landsteiner}, \citenamefont {Romeu},\ and\
  \citenamefont {Saura-Bastida}}]{Garcia-Farina:2024pdd}%
  \BibitemOpen
  \bibfield  {author} {\bibinfo {author} {\bibfnamefont {D.}~\bibnamefont
  {Garcia-Fari\~na}}, \bibinfo {author} {\bibfnamefont {K.}~\bibnamefont
  {Landsteiner}}, \bibinfo {author} {\bibfnamefont {P.~G.}\ \bibnamefont
  {Romeu}}, \ and\ \bibinfo {author} {\bibfnamefont {P.}~\bibnamefont
  {Saura-Bastida}},\ }\href@noop {} {\  (\bibinfo {year} {2024})},\ \Eprint
  {http://arxiv.org/abs/2407.06104} {arXiv:2407.06104 [hep-th]} \BibitemShut
  {NoStop}%
\bibitem [{\citenamefont {Are\'an}\ \emph {et~al.}(2023)\citenamefont
  {Are\'an}, \citenamefont {Fari\~na},\ and\ \citenamefont
  {Landsteiner}}]{Arean:2023ejh}%
  \BibitemOpen
  \bibfield  {author} {\bibinfo {author} {\bibfnamefont {D.}~\bibnamefont
  {Are\'an}}, \bibinfo {author} {\bibfnamefont {D.~G.}\ \bibnamefont
  {Fari\~na}}, \ and\ \bibinfo {author} {\bibfnamefont {K.}~\bibnamefont
  {Landsteiner}},\ }\href {\doibase 10.1007/JHEP12(2023)187} {\bibfield
  {journal} {\bibinfo  {journal} {JHEP}\ }\textbf {\bibinfo {volume} {12}},\
  \bibinfo {pages} {187} (\bibinfo {year} {2023})},\ \Eprint
  {http://arxiv.org/abs/2307.08751} {arXiv:2307.08751 [hep-th]} \BibitemShut
  {NoStop}%
\bibitem [{\citenamefont {Boyanov}\ \emph {et~al.}(2024)\citenamefont
  {Boyanov}, \citenamefont {Cardoso}, \citenamefont {Destounis}, \citenamefont
  {Jaramillo},\ and\ \citenamefont {Panosso~Macedo}}]{Boyanov:2023qqf}%
  \BibitemOpen
  \bibfield  {author} {\bibinfo {author} {\bibfnamefont {V.}~\bibnamefont
  {Boyanov}}, \bibinfo {author} {\bibfnamefont {V.}~\bibnamefont {Cardoso}},
  \bibinfo {author} {\bibfnamefont {K.}~\bibnamefont {Destounis}}, \bibinfo
  {author} {\bibfnamefont {J.~L.}\ \bibnamefont {Jaramillo}}, \ and\ \bibinfo
  {author} {\bibfnamefont {R.}~\bibnamefont {Panosso~Macedo}},\ }\href
  {\doibase 10.1103/PhysRevD.109.064068} {\bibfield  {journal} {\bibinfo
  {journal} {Phys. Rev. D}\ }\textbf {\bibinfo {volume} {109}},\ \bibinfo
  {pages} {064068} (\bibinfo {year} {2024})},\ \Eprint
  {http://arxiv.org/abs/2312.11998} {arXiv:2312.11998 [gr-qc]} \BibitemShut
  {NoStop}%
\bibitem [{\citenamefont {Cownden}\ \emph {et~al.}(2024)\citenamefont
  {Cownden}, \citenamefont {Pantelidou},\ and\ \citenamefont
  {Zilh\~ao}}]{Cownden:2023dam}%
  \BibitemOpen
  \bibfield  {author} {\bibinfo {author} {\bibfnamefont {B.}~\bibnamefont
  {Cownden}}, \bibinfo {author} {\bibfnamefont {C.}~\bibnamefont {Pantelidou}},
  \ and\ \bibinfo {author} {\bibfnamefont {M.}~\bibnamefont {Zilh\~ao}},\
  }\href {\doibase 10.1007/JHEP05(2024)202} {\bibfield  {journal} {\bibinfo
  {journal} {JHEP}\ }\textbf {\bibinfo {volume} {05}},\ \bibinfo {pages} {202}
  (\bibinfo {year} {2024})},\ \Eprint {http://arxiv.org/abs/2312.08352}
  {arXiv:2312.08352 [gr-qc]} \BibitemShut {NoStop}%
\bibitem [{\citenamefont {Sarkar}\ \emph {et~al.}(2023)\citenamefont {Sarkar},
  \citenamefont {Rahman},\ and\ \citenamefont {Chakraborty}}]{Sarkar:2023rhp}%
  \BibitemOpen
  \bibfield  {author} {\bibinfo {author} {\bibfnamefont {S.}~\bibnamefont
  {Sarkar}}, \bibinfo {author} {\bibfnamefont {M.}~\bibnamefont {Rahman}}, \
  and\ \bibinfo {author} {\bibfnamefont {S.}~\bibnamefont {Chakraborty}},\
  }\href {\doibase 10.1103/PhysRevD.108.104002} {\bibfield  {journal} {\bibinfo
   {journal} {Phys. Rev. D}\ }\textbf {\bibinfo {volume} {108}},\ \bibinfo
  {pages} {104002} (\bibinfo {year} {2023})},\ \Eprint
  {http://arxiv.org/abs/2304.06829} {arXiv:2304.06829 [gr-qc]} \BibitemShut
  {NoStop}%
\bibitem [{\citenamefont {Destounis}\ \emph {et~al.}(2024)\citenamefont
  {Destounis}, \citenamefont {Boyanov},\ and\ \citenamefont
  {Panosso~Macedo}}]{Destounis:2023nmb}%
  \BibitemOpen
  \bibfield  {author} {\bibinfo {author} {\bibfnamefont {K.}~\bibnamefont
  {Destounis}}, \bibinfo {author} {\bibfnamefont {V.}~\bibnamefont {Boyanov}},
  \ and\ \bibinfo {author} {\bibfnamefont {R.}~\bibnamefont {Panosso~Macedo}},\
  }\href {\doibase 10.1103/PhysRevD.109.044023} {\bibfield  {journal} {\bibinfo
   {journal} {Phys. Rev. D}\ }\textbf {\bibinfo {volume} {109}},\ \bibinfo
  {pages} {044023} (\bibinfo {year} {2024})},\ \Eprint
  {http://arxiv.org/abs/2312.11630} {arXiv:2312.11630 [gr-qc]} \BibitemShut
  {NoStop}%
\bibitem [{\citenamefont {Luo}(2024)}]{Luo:2024dxl}%
  \BibitemOpen
  \bibfield  {author} {\bibinfo {author} {\bibfnamefont {S.}~\bibnamefont
  {Luo}},\ }\href {\doibase 10.1103/PhysRevD.110.084071} {\bibfield  {journal}
  {\bibinfo  {journal} {Phys. Rev. D}\ }\textbf {\bibinfo {volume} {110}},\
  \bibinfo {pages} {084071} (\bibinfo {year} {2024})},\ \Eprint
  {http://arxiv.org/abs/2408.08139} {arXiv:2408.08139 [gr-qc]} \BibitemShut
  {NoStop}%
\bibitem [{\citenamefont {Warnick}(2024)}]{Warnick:2024usx}%
  \BibitemOpen
  \bibfield  {author} {\bibinfo {author} {\bibfnamefont {C.}~\bibnamefont
  {Warnick}},\ }\href@noop {} {\  (\bibinfo {year} {2024})},\ \Eprint
  {http://arxiv.org/abs/2407.19850} {arXiv:2407.19850 [gr-qc]} \BibitemShut
  {NoStop}%
\bibitem [{\citenamefont {Boyanov}\ \emph {et~al.}(2023)\citenamefont
  {Boyanov}, \citenamefont {Destounis}, \citenamefont {Panosso~Macedo},
  \citenamefont {Cardoso},\ and\ \citenamefont {Jaramillo}}]{Boyanov:2022ark}%
  \BibitemOpen
  \bibfield  {author} {\bibinfo {author} {\bibfnamefont {V.}~\bibnamefont
  {Boyanov}}, \bibinfo {author} {\bibfnamefont {K.}~\bibnamefont {Destounis}},
  \bibinfo {author} {\bibfnamefont {R.}~\bibnamefont {Panosso~Macedo}},
  \bibinfo {author} {\bibfnamefont {V.}~\bibnamefont {Cardoso}}, \ and\
  \bibinfo {author} {\bibfnamefont {J.~L.}\ \bibnamefont {Jaramillo}},\ }\href
  {\doibase 10.1103/PhysRevD.107.064012} {\bibfield  {journal} {\bibinfo
  {journal} {Phys. Rev. D}\ }\textbf {\bibinfo {volume} {107}},\ \bibinfo
  {pages} {064012} (\bibinfo {year} {2023})},\ \Eprint
  {http://arxiv.org/abs/2209.12950} {arXiv:2209.12950 [gr-qc]} \BibitemShut
  {NoStop}%
\bibitem [{\citenamefont {Carballo}\ and\ \citenamefont
  {Withers}(2024)}]{Carballo:2024kbk}%
  \BibitemOpen
  \bibfield  {author} {\bibinfo {author} {\bibfnamefont {J.}~\bibnamefont
  {Carballo}}\ and\ \bibinfo {author} {\bibfnamefont {B.}~\bibnamefont
  {Withers}},\ }\href {\doibase 10.1007/JHEP10(2024)084} {\bibfield  {journal}
  {\bibinfo  {journal} {JHEP}\ }\textbf {\bibinfo {volume} {10}},\ \bibinfo
  {pages} {084} (\bibinfo {year} {2024})},\ \Eprint
  {http://arxiv.org/abs/2406.06685} {arXiv:2406.06685 [hep-th]} \BibitemShut
  {NoStop}%
\bibitem [{\citenamefont {Jaramillo}(2022)}]{Jaramillo:2022kuv}%
  \BibitemOpen
  \bibfield  {author} {\bibinfo {author} {\bibfnamefont {J.~L.}\ \bibnamefont
  {Jaramillo}},\ }\href {\doibase 10.1088/1361-6382/ac8ddc} {\bibfield
  {journal} {\bibinfo  {journal} {Class. Quant. Grav.}\ }\textbf {\bibinfo
  {volume} {39}},\ \bibinfo {pages} {217002} (\bibinfo {year} {2022})},\
  \Eprint {http://arxiv.org/abs/2206.08025} {arXiv:2206.08025 [gr-qc]}
  \BibitemShut {NoStop}%
\bibitem [{\citenamefont {Chen}\ \emph {et~al.}(2024)\citenamefont {Chen},
  \citenamefont {Wu},\ and\ \citenamefont {Guo}}]{Chen:2024mon}%
  \BibitemOpen
  \bibfield  {author} {\bibinfo {author} {\bibfnamefont {J.-N.}\ \bibnamefont
  {Chen}}, \bibinfo {author} {\bibfnamefont {L.-B.}\ \bibnamefont {Wu}}, \ and\
  \bibinfo {author} {\bibfnamefont {Z.-K.}\ \bibnamefont {Guo}},\ }\href
  {\doibase 10.1088/1361-6382/ad89a1} {\bibfield  {journal} {\bibinfo
  {journal} {Class. Quant. Grav.}\ }\textbf {\bibinfo {volume} {41}},\ \bibinfo
  {pages} {235015} (\bibinfo {year} {2024})},\ \Eprint
  {http://arxiv.org/abs/2407.03907} {arXiv:2407.03907 [gr-qc]} \BibitemShut
  {NoStop}%
\bibitem [{\citenamefont {Besson}\ and\ \citenamefont
  {Jaramillo}(2024)}]{Besson:2024adi}%
  \BibitemOpen
  \bibfield  {author} {\bibinfo {author} {\bibfnamefont {J.}~\bibnamefont
  {Besson}}\ and\ \bibinfo {author} {\bibfnamefont {J.~L.}\ \bibnamefont
  {Jaramillo}},\ }\href@noop {} {\  (\bibinfo {year} {2024})},\ \Eprint
  {http://arxiv.org/abs/2412.02793} {arXiv:2412.02793 [gr-qc]} \BibitemShut
  {NoStop}%
\bibitem [{\citenamefont {Panosso~Macedo}\ and\ \citenamefont
  {Zenginoglu}(2024)}]{PanossoMacedo:2024nkw}%
  \BibitemOpen
  \bibfield  {author} {\bibinfo {author} {\bibfnamefont {R.}~\bibnamefont
  {Panosso~Macedo}}\ and\ \bibinfo {author} {\bibfnamefont {A.}~\bibnamefont
  {Zenginoglu}},\ }\href@noop {} {\  (\bibinfo {year} {2024})},\ \Eprint
  {http://arxiv.org/abs/2409.11478} {arXiv:2409.11478 [gr-qc]} \BibitemShut
  {NoStop}%
\bibitem [{\citenamefont {Zenginoglu}(2008)}]{Zenginoglu:2007jw}%
  \BibitemOpen
  \bibfield  {author} {\bibinfo {author} {\bibfnamefont {A.}~\bibnamefont
  {Zenginoglu}},\ }\href {\doibase 10.1088/0264-9381/25/14/145002} {\bibfield
  {journal} {\bibinfo  {journal} {Class. Quant. Grav.}\ }\textbf {\bibinfo
  {volume} {25}},\ \bibinfo {pages} {145002} (\bibinfo {year} {2008})},\
  \Eprint {http://arxiv.org/abs/0712.4333} {arXiv:0712.4333 [gr-qc]}
  \BibitemShut {NoStop}%
\bibitem [{\citenamefont {Zenginoglu}(2011)}]{Zenginoglu:2011jz}%
  \BibitemOpen
  \bibfield  {author} {\bibinfo {author} {\bibfnamefont {A.}~\bibnamefont
  {Zenginoglu}},\ }\href {\doibase 10.1103/PhysRevD.83.127502} {\bibfield
  {journal} {\bibinfo  {journal} {Phys. Rev. D}\ }\textbf {\bibinfo {volume}
  {83}},\ \bibinfo {pages} {127502} (\bibinfo {year} {2011})},\ \Eprint
  {http://arxiv.org/abs/1102.2451} {arXiv:1102.2451 [gr-qc]} \BibitemShut
  {NoStop}%
\bibitem [{\citenamefont {Panosso~Macedo}(2024)}]{PanossoMacedo:2023qzp}%
  \BibitemOpen
  \bibfield  {author} {\bibinfo {author} {\bibfnamefont {R.}~\bibnamefont
  {Panosso~Macedo}},\ }\href {\doibase 10.1098/rsta.2023.0046} {\bibfield
  {journal} {\bibinfo  {journal} {Phil. Trans. Roy. Soc. Lond. A}\ }\textbf
  {\bibinfo {volume} {382}},\ \bibinfo {pages} {20230046} (\bibinfo {year}
  {2024})},\ \Eprint {http://arxiv.org/abs/2307.15735} {arXiv:2307.15735
  [gr-qc]} \BibitemShut {NoStop}%
\bibitem [{\citenamefont {Zengino\u{g}lu}(2024)}]{Zenginoglu:2024bzs}%
  \BibitemOpen
  \bibfield  {author} {\bibinfo {author} {\bibfnamefont {A.}~\bibnamefont
  {Zengino\u{g}lu}},\ }\href {\doibase 10.1119/5.0214271} {\bibfield  {journal}
  {\bibinfo  {journal} {Am. J. Phys.}\ }\textbf {\bibinfo {volume} {92}},\
  \bibinfo {pages} {965} (\bibinfo {year} {2024})},\ \Eprint
  {http://arxiv.org/abs/2404.01528} {arXiv:2404.01528 [gr-qc]} \BibitemShut
  {NoStop}%
\bibitem [{\citenamefont {Panosso~Macedo}(2020)}]{PanossoMacedo:2019npm}%
  \BibitemOpen
  \bibfield  {author} {\bibinfo {author} {\bibfnamefont {R.}~\bibnamefont
  {Panosso~Macedo}},\ }\href {\doibase 10.1088/1361-6382/ab6e3e} {\bibfield
  {journal} {\bibinfo  {journal} {Class. Quant. Grav.}\ }\textbf {\bibinfo
  {volume} {37}},\ \bibinfo {pages} {065019} (\bibinfo {year} {2020})},\
  \Eprint {http://arxiv.org/abs/1910.13452} {arXiv:1910.13452 [gr-qc]}
  \BibitemShut {NoStop}%
\bibitem [{\citenamefont {Minucci}\ and\ \citenamefont
  {Panosso~Macedo}(2024)}]{Minucci:2024qrn}%
  \BibitemOpen
  \bibfield  {author} {\bibinfo {author} {\bibfnamefont {M.}~\bibnamefont
  {Minucci}}\ and\ \bibinfo {author} {\bibfnamefont {R.}~\bibnamefont
  {Panosso~Macedo}},\ }\href@noop {} {\  (\bibinfo {year} {2024})},\ \Eprint
  {http://arxiv.org/abs/2411.19740} {arXiv:2411.19740 [gr-qc]} \BibitemShut
  {NoStop}%
\bibitem [{\citenamefont {Chen}\ and\ \citenamefont
  {Jing}(2023)}]{Chen:2023ese}%
  \BibitemOpen
  \bibfield  {author} {\bibinfo {author} {\bibfnamefont {C.}~\bibnamefont
  {Chen}}\ and\ \bibinfo {author} {\bibfnamefont {J.}~\bibnamefont {Jing}},\
  }\href {\doibase 10.1088/1475-7516/2023/11/070} {\bibfield  {journal}
  {\bibinfo  {journal} {JCAP}\ }\textbf {\bibinfo {volume} {11}},\ \bibinfo
  {pages} {070} (\bibinfo {year} {2023})},\ \Eprint
  {http://arxiv.org/abs/2307.14616} {arXiv:2307.14616 [gr-qc]} \BibitemShut
  {NoStop}%
\bibitem [{\citenamefont {Bl\'azquez-Salcedo}\ \emph
  {et~al.}(2024{\natexlab{a}})\citenamefont {Bl\'azquez-Salcedo}, \citenamefont
  {Khoo}, \citenamefont {Kunz},\ and\ \citenamefont
  {Gonz\'alez-Romero}}]{Blazquez-Salcedo:2023hwg}%
  \BibitemOpen
  \bibfield  {author} {\bibinfo {author} {\bibfnamefont {J.~L.}\ \bibnamefont
  {Bl\'azquez-Salcedo}}, \bibinfo {author} {\bibfnamefont {F.~S.}\ \bibnamefont
  {Khoo}}, \bibinfo {author} {\bibfnamefont {J.}~\bibnamefont {Kunz}}, \ and\
  \bibinfo {author} {\bibfnamefont {L.~M.}\ \bibnamefont {Gonz\'alez-Romero}},\
  }\href {\doibase 10.1103/PhysRevD.109.064028} {\bibfield  {journal} {\bibinfo
   {journal} {Phys. Rev. D}\ }\textbf {\bibinfo {volume} {109}},\ \bibinfo
  {pages} {064028} (\bibinfo {year} {2024}{\natexlab{a}})},\ \Eprint
  {http://arxiv.org/abs/2312.10754} {arXiv:2312.10754 [gr-qc]} \BibitemShut
  {NoStop}%
\bibitem [{\citenamefont {Chung}\ \emph {et~al.}(2024)\citenamefont {Chung},
  \citenamefont {Wagle},\ and\ \citenamefont {Yunes}}]{Chung:2023wkd}%
  \BibitemOpen
  \bibfield  {author} {\bibinfo {author} {\bibfnamefont {A.~K.-W.}\
  \bibnamefont {Chung}}, \bibinfo {author} {\bibfnamefont {P.}~\bibnamefont
  {Wagle}}, \ and\ \bibinfo {author} {\bibfnamefont {N.}~\bibnamefont
  {Yunes}},\ }\href {\doibase 10.1103/PhysRevD.109.044072} {\bibfield
  {journal} {\bibinfo  {journal} {Phys. Rev. D}\ }\textbf {\bibinfo {volume}
  {109}},\ \bibinfo {pages} {044072} (\bibinfo {year} {2024})},\ \Eprint
  {http://arxiv.org/abs/2312.08435} {arXiv:2312.08435 [gr-qc]} \BibitemShut
  {NoStop}%
\bibitem [{\citenamefont {Khoo}\ \emph {et~al.}(2024)\citenamefont {Khoo},
  \citenamefont {Azad}, \citenamefont {Bl\'azquez-Salcedo}, \citenamefont
  {Gonz\'alez-Romero}, \citenamefont {Kleihaus}, \citenamefont {Kunz},\ and\
  \citenamefont {Navarro-L\'erida}}]{Khoo:2024yeh}%
  \BibitemOpen
  \bibfield  {author} {\bibinfo {author} {\bibfnamefont {F.~S.}\ \bibnamefont
  {Khoo}}, \bibinfo {author} {\bibfnamefont {B.}~\bibnamefont {Azad}}, \bibinfo
  {author} {\bibfnamefont {J.~L.}\ \bibnamefont {Bl\'azquez-Salcedo}}, \bibinfo
  {author} {\bibfnamefont {L.~M.}\ \bibnamefont {Gonz\'alez-Romero}}, \bibinfo
  {author} {\bibfnamefont {B.}~\bibnamefont {Kleihaus}}, \bibinfo {author}
  {\bibfnamefont {J.}~\bibnamefont {Kunz}}, \ and\ \bibinfo {author}
  {\bibfnamefont {F.}~\bibnamefont {Navarro-L\'erida}},\ }\href {\doibase
  10.1103/PhysRevD.109.084013} {\bibfield  {journal} {\bibinfo  {journal}
  {Phys. Rev. D}\ }\textbf {\bibinfo {volume} {109}},\ \bibinfo {pages}
  {084013} (\bibinfo {year} {2024})},\ \Eprint
  {http://arxiv.org/abs/2401.02898} {arXiv:2401.02898 [gr-qc]} \BibitemShut
  {NoStop}%
\bibitem [{\citenamefont {Chung}\ and\ \citenamefont
  {Yunes}(2024{\natexlab{a}})}]{Chung:2024ira}%
  \BibitemOpen
  \bibfield  {author} {\bibinfo {author} {\bibfnamefont {A.~K.-W.}\
  \bibnamefont {Chung}}\ and\ \bibinfo {author} {\bibfnamefont
  {N.}~\bibnamefont {Yunes}},\ }\href {\doibase 10.1103/PhysRevLett.133.181401}
  {\bibfield  {journal} {\bibinfo  {journal} {Phys. Rev. Lett.}\ }\textbf
  {\bibinfo {volume} {133}},\ \bibinfo {pages} {181401} (\bibinfo {year}
  {2024}{\natexlab{a}})},\ \Eprint {http://arxiv.org/abs/2405.12280}
  {arXiv:2405.12280 [gr-qc]} \BibitemShut {NoStop}%
\bibitem [{\citenamefont {Chung}\ and\ \citenamefont
  {Yunes}(2024{\natexlab{b}})}]{Chung:2024vaf}%
  \BibitemOpen
  \bibfield  {author} {\bibinfo {author} {\bibfnamefont {A.~K.-W.}\
  \bibnamefont {Chung}}\ and\ \bibinfo {author} {\bibfnamefont
  {N.}~\bibnamefont {Yunes}},\ }\href {\doibase 10.1103/PhysRevD.110.064019}
  {\bibfield  {journal} {\bibinfo  {journal} {Phys. Rev. D}\ }\textbf {\bibinfo
  {volume} {110}},\ \bibinfo {pages} {064019} (\bibinfo {year}
  {2024}{\natexlab{b}})},\ \Eprint {http://arxiv.org/abs/2406.11986}
  {arXiv:2406.11986 [gr-qc]} \BibitemShut {NoStop}%
\bibitem [{\citenamefont {Bl\'azquez-Salcedo}\ \emph
  {et~al.}(2024{\natexlab{b}})\citenamefont {Bl\'azquez-Salcedo}, \citenamefont
  {Khoo}, \citenamefont {Kleihaus},\ and\ \citenamefont
  {Kunz}}]{Blazquez-Salcedo:2024oek}%
  \BibitemOpen
  \bibfield  {author} {\bibinfo {author} {\bibfnamefont {J.~L.}\ \bibnamefont
  {Bl\'azquez-Salcedo}}, \bibinfo {author} {\bibfnamefont {F.~S.}\ \bibnamefont
  {Khoo}}, \bibinfo {author} {\bibfnamefont {B.}~\bibnamefont {Kleihaus}}, \
  and\ \bibinfo {author} {\bibfnamefont {J.}~\bibnamefont {Kunz}},\ }\href@noop
  {} {\  (\bibinfo {year} {2024}{\natexlab{b}})},\ \Eprint
  {http://arxiv.org/abs/2407.20760} {arXiv:2407.20760 [gr-qc]} \BibitemShut
  {NoStop}%
\bibitem [{\citenamefont {Bl\'azquez-Salcedo}\ \emph
  {et~al.}(2024{\natexlab{c}})\citenamefont {Bl\'azquez-Salcedo}, \citenamefont
  {Khoo}, \citenamefont {Kleihaus},\ and\ \citenamefont
  {Kunz}}]{Blazquez-Salcedo:2024dur}%
  \BibitemOpen
  \bibfield  {author} {\bibinfo {author} {\bibfnamefont {J.~L.}\ \bibnamefont
  {Bl\'azquez-Salcedo}}, \bibinfo {author} {\bibfnamefont {F.~S.}\ \bibnamefont
  {Khoo}}, \bibinfo {author} {\bibfnamefont {B.}~\bibnamefont {Kleihaus}}, \
  and\ \bibinfo {author} {\bibfnamefont {J.}~\bibnamefont {Kunz}},\ }\href@noop
  {} {\  (\bibinfo {year} {2024}{\natexlab{c}})},\ \Eprint
  {http://arxiv.org/abs/2412.17073} {arXiv:2412.17073 [gr-qc]} \BibitemShut
  {NoStop}%
\bibitem [{\citenamefont {Ripley}(2022)}]{Ripley:2022ypi}%
  \BibitemOpen
  \bibfield  {author} {\bibinfo {author} {\bibfnamefont {J.~L.}\ \bibnamefont
  {Ripley}},\ }\href {\doibase 10.1088/1361-6382/ac776d} {\bibfield  {journal}
  {\bibinfo  {journal} {Class. Quant. Grav.}\ }\textbf {\bibinfo {volume}
  {39}},\ \bibinfo {pages} {145009} (\bibinfo {year} {2022})},\ \Eprint
  {http://arxiv.org/abs/2202.03837} {arXiv:2202.03837 [gr-qc]} \BibitemShut
  {NoStop}%
\bibitem [{\citenamefont {Warnick}(2015)}]{Warnick:2013hba}%
  \BibitemOpen
  \bibfield  {author} {\bibinfo {author} {\bibfnamefont {C.~M.}\ \bibnamefont
  {Warnick}},\ }\href {\doibase 10.1007/s00220-014-2171-1} {\bibfield
  {journal} {\bibinfo  {journal} {Commun. Math. Phys.}\ }\textbf {\bibinfo
  {volume} {333}},\ \bibinfo {pages} {959} (\bibinfo {year} {2015})},\ \Eprint
  {http://arxiv.org/abs/1306.5760} {arXiv:1306.5760 [gr-qc]} \BibitemShut
  {NoStop}%
\bibitem [{\citenamefont {Gasperin}\ and\ \citenamefont
  {Jaramillo}(2022)}]{Gasperin:2021kfv}%
  \BibitemOpen
  \bibfield  {author} {\bibinfo {author} {\bibfnamefont {E.}~\bibnamefont
  {Gasperin}}\ and\ \bibinfo {author} {\bibfnamefont {J.~L.}\ \bibnamefont
  {Jaramillo}},\ }\href {\doibase 10.1088/1361-6382/ac5054} {\bibfield
  {journal} {\bibinfo  {journal} {Class. Quant. Grav.}\ }\textbf {\bibinfo
  {volume} {39}},\ \bibinfo {pages} {115010} (\bibinfo {year} {2022})},\
  \Eprint {http://arxiv.org/abs/2107.12865} {arXiv:2107.12865 [gr-qc]}
  \BibitemShut {NoStop}%
\bibitem [{\citenamefont {Bini}\ \emph {et~al.}(2002)\citenamefont {Bini},
  \citenamefont {Cherubini}, \citenamefont {Jantzen},\ and\ \citenamefont
  {Ruffini}}]{Bini:2002jx}%
  \BibitemOpen
  \bibfield  {author} {\bibinfo {author} {\bibfnamefont {D.}~\bibnamefont
  {Bini}}, \bibinfo {author} {\bibfnamefont {C.}~\bibnamefont {Cherubini}},
  \bibinfo {author} {\bibfnamefont {R.~T.}\ \bibnamefont {Jantzen}}, \ and\
  \bibinfo {author} {\bibfnamefont {R.~J.}\ \bibnamefont {Ruffini}},\ }\href
  {\doibase 10.1143/PTP.107.967} {\bibfield  {journal} {\bibinfo  {journal}
  {Prog. Theor. Phys.}\ }\textbf {\bibinfo {volume} {107}},\ \bibinfo {pages}
  {967} (\bibinfo {year} {2002})},\ \Eprint
  {http://arxiv.org/abs/gr-qc/0203069} {arXiv:gr-qc/0203069} \BibitemShut
  {NoStop}%
\bibitem [{\citenamefont {Pound}(2015)}]{Pound:2015wva}%
  \BibitemOpen
  \bibfield  {author} {\bibinfo {author} {\bibfnamefont {A.}~\bibnamefont
  {Pound}},\ }\href {\doibase 10.1103/PhysRevD.92.104047} {\bibfield  {journal}
  {\bibinfo  {journal} {Phys. Rev. D}\ }\textbf {\bibinfo {volume} {92}},\
  \bibinfo {pages} {104047} (\bibinfo {year} {2015})},\ \Eprint
  {http://arxiv.org/abs/1510.05172} {arXiv:1510.05172 [gr-qc]} \BibitemShut
  {NoStop}%
\bibitem [{\citenamefont {Xiong}\ and\ \citenamefont
  {Li}(2024)}]{Xiong:2024urw}%
  \BibitemOpen
  \bibfield  {author} {\bibinfo {author} {\bibfnamefont {W.}~\bibnamefont
  {Xiong}}\ and\ \bibinfo {author} {\bibfnamefont {P.-C.}\ \bibnamefont {Li}},\
  }\href@noop {} {\  (\bibinfo {year} {2024})},\ \Eprint
  {http://arxiv.org/abs/2411.19069} {arXiv:2411.19069 [gr-qc]} \BibitemShut
  {NoStop}%
\bibitem [{\citenamefont {Leaver}(1985)}]{Leaver:1985ax}%
  \BibitemOpen
  \bibfield  {author} {\bibinfo {author} {\bibfnamefont {E.~W.}\ \bibnamefont
  {Leaver}},\ }\href {\doibase 10.1098/rspa.1985.0119} {\bibfield  {journal}
  {\bibinfo  {journal} {Proc. Roy. Soc. Lond. A}\ }\textbf {\bibinfo {volume}
  {402}},\ \bibinfo {pages} {285} (\bibinfo {year} {1985})}\BibitemShut
  {NoStop}%
\bibitem [{\citenamefont {Berti}\ \emph {et~al.}(2009)\citenamefont {Berti},
  \citenamefont {Cardoso},\ and\ \citenamefont {Starinets}}]{Berti:2009kk}%
  \BibitemOpen
  \bibfield  {author} {\bibinfo {author} {\bibfnamefont {E.}~\bibnamefont
  {Berti}}, \bibinfo {author} {\bibfnamefont {V.}~\bibnamefont {Cardoso}}, \
  and\ \bibinfo {author} {\bibfnamefont {A.~O.}\ \bibnamefont {Starinets}},\
  }\href {\doibase 10.1088/0264-9381/26/16/163001} {\bibfield  {journal}
  {\bibinfo  {journal} {Class. Quant. Grav.}\ }\textbf {\bibinfo {volume}
  {26}},\ \bibinfo {pages} {163001} (\bibinfo {year} {2009})},\ \Eprint
  {http://arxiv.org/abs/0905.2975} {arXiv:0905.2975 [gr-qc]} \BibitemShut
  {NoStop}%
\bibitem [{\citenamefont {T\'oth}(2018)}]{Toth:2018qrx}%
  \BibitemOpen
  \bibfield  {author} {\bibinfo {author} {\bibfnamefont {G.~a.~Z.}\
  \bibnamefont {T\'oth}},\ }\href {\doibase 10.1088/1361-6382/aad712}
  {\bibfield  {journal} {\bibinfo  {journal} {Class. Quant. Grav.}\ }\textbf
  {\bibinfo {volume} {35}},\ \bibinfo {pages} {185009} (\bibinfo {year}
  {2018})},\ \Eprint {http://arxiv.org/abs/1801.04710} {arXiv:1801.04710
  [gr-qc]} \BibitemShut {NoStop}%
\bibitem [{\citenamefont {Csuk\'as}\ \emph {et~al.}(2019)\citenamefont
  {Csuk\'as}, \citenamefont {R\'acz},\ and\ \citenamefont
  {T\'oth}}]{Csukas:2019kcb}%
  \BibitemOpen
  \bibfield  {author} {\bibinfo {author} {\bibfnamefont {K.}~\bibnamefont
  {Csuk\'as}}, \bibinfo {author} {\bibfnamefont {I.}~\bibnamefont {R\'acz}}, \
  and\ \bibinfo {author} {\bibfnamefont {G.~Z.}\ \bibnamefont {T\'oth}},\
  }\href {\doibase 10.1103/PhysRevD.100.104025} {\bibfield  {journal} {\bibinfo
   {journal} {Phys. Rev. D}\ }\textbf {\bibinfo {volume} {100}},\ \bibinfo
  {pages} {104025} (\bibinfo {year} {2019})},\ \Eprint
  {http://arxiv.org/abs/1905.09082} {arXiv:1905.09082 [gr-qc]} \BibitemShut
  {NoStop}%
\bibitem [{\citenamefont {Motohashi}(2024)}]{Motohashi:2024fwt}%
  \BibitemOpen
  \bibfield  {author} {\bibinfo {author} {\bibfnamefont {H.}~\bibnamefont
  {Motohashi}},\ }\href@noop {} {\  (\bibinfo {year} {2024})},\ \Eprint
  {http://arxiv.org/abs/2407.15191} {arXiv:2407.15191 [gr-qc]} \BibitemShut
  {NoStop}%
\bibitem [{\citenamefont {London}(2023)}]{London:2023aeo}%
  \BibitemOpen
  \bibfield  {author} {\bibinfo {author} {\bibfnamefont {L.~T.}\ \bibnamefont
  {London}},\ }\href@noop {} {\  (\bibinfo {year} {2023})},\ \Eprint
  {http://arxiv.org/abs/2312.17678} {arXiv:2312.17678 [gr-qc]} \BibitemShut
  {NoStop}%
\bibitem [{\citenamefont {London}\ and\ \citenamefont
  {Gurevich}(2023)}]{London:2023idh}%
  \BibitemOpen
  \bibfield  {author} {\bibinfo {author} {\bibfnamefont {L.}~\bibnamefont
  {London}}\ and\ \bibinfo {author} {\bibfnamefont {M.}~\bibnamefont
  {Gurevich}},\ }\href@noop {} {\  (\bibinfo {year} {2023})},\ \Eprint
  {http://arxiv.org/abs/2312.17680} {arXiv:2312.17680 [gr-qc]} \BibitemShut
  {NoStop}%
\bibitem [{\citenamefont {Green}\ \emph {et~al.}(2023)\citenamefont {Green},
  \citenamefont {Hollands}, \citenamefont {Sberna}, \citenamefont {Toomani},\
  and\ \citenamefont {Zimmerman}}]{Green:2022htq}%
  \BibitemOpen
  \bibfield  {author} {\bibinfo {author} {\bibfnamefont {S.~R.}\ \bibnamefont
  {Green}}, \bibinfo {author} {\bibfnamefont {S.}~\bibnamefont {Hollands}},
  \bibinfo {author} {\bibfnamefont {L.}~\bibnamefont {Sberna}}, \bibinfo
  {author} {\bibfnamefont {V.}~\bibnamefont {Toomani}}, \ and\ \bibinfo
  {author} {\bibfnamefont {P.}~\bibnamefont {Zimmerman}},\ }\href {\doibase
  10.1103/PhysRevD.107.064030} {\bibfield  {journal} {\bibinfo  {journal}
  {Phys. Rev. D}\ }\textbf {\bibinfo {volume} {107}},\ \bibinfo {pages}
  {064030} (\bibinfo {year} {2023})},\ \Eprint
  {http://arxiv.org/abs/2210.15935} {arXiv:2210.15935 [gr-qc]} \BibitemShut
  {NoStop}%
\bibitem [{\citenamefont {Trefethen}(1999)}]{trefethen_1999}%
  \BibitemOpen
  \bibfield  {author} {\bibinfo {author} {\bibfnamefont {L.~N.}\ \bibnamefont
  {Trefethen}},\ }\href {\doibase 10.1017/S0962492900002932} {\bibfield
  {journal} {\bibinfo  {journal} {Acta Numerica}\ }\textbf {\bibinfo {volume}
  {8}},\ \bibinfo {pages} {247–295} (\bibinfo {year} {1999})}\BibitemShut
  {NoStop}%
\bibitem [{\citenamefont {Ashida}\ \emph {et~al.}(2021)\citenamefont {Ashida},
  \citenamefont {Gong},\ and\ \citenamefont {Ueda}}]{Ashida:2020dkc}%
  \BibitemOpen
  \bibfield  {author} {\bibinfo {author} {\bibfnamefont {Y.}~\bibnamefont
  {Ashida}}, \bibinfo {author} {\bibfnamefont {Z.}~\bibnamefont {Gong}}, \ and\
  \bibinfo {author} {\bibfnamefont {M.}~\bibnamefont {Ueda}},\ }\href {\doibase
  10.1080/00018732.2021.1876991} {\bibfield  {journal} {\bibinfo  {journal}
  {Adv. Phys.}\ }\textbf {\bibinfo {volume} {69}},\ \bibinfo {pages} {249}
  (\bibinfo {year} {2021})},\ \Eprint {http://arxiv.org/abs/2006.01837}
  {arXiv:2006.01837 [cond-mat.mes-hall]} \BibitemShut {NoStop}%
\bibitem [{\citenamefont {Krejcirik}\ \emph {et~al.}(2015)\citenamefont
  {Krejcirik}, \citenamefont {Siegl}, \citenamefont {Tater},\ and\
  \citenamefont {Viola}}]{Krejcirik:2014kaa}%
  \BibitemOpen
  \bibfield  {author} {\bibinfo {author} {\bibfnamefont {D.}~\bibnamefont
  {Krejcirik}}, \bibinfo {author} {\bibfnamefont {P.}~\bibnamefont {Siegl}},
  \bibinfo {author} {\bibfnamefont {M.}~\bibnamefont {Tater}}, \ and\ \bibinfo
  {author} {\bibfnamefont {J.}~\bibnamefont {Viola}},\ }\href {\doibase
  10.1063/1.4934378} {\bibfield  {journal} {\bibinfo  {journal} {J. Math.
  Phys.}\ }\textbf {\bibinfo {volume} {56}},\ \bibinfo {pages} {103513}
  (\bibinfo {year} {2015})},\ \Eprint {http://arxiv.org/abs/1402.1082}
  {arXiv:1402.1082 [math.SP]} \BibitemShut {NoStop}%
\bibitem [{\citenamefont {Markakis}\ \emph {et~al.}(2019)\citenamefont
  {Markakis}, \citenamefont {O'Boyle}, \citenamefont {Glennon}, \citenamefont
  {Tran}, \citenamefont {Brubeck}, \citenamefont {Haas}, \citenamefont
  {Schive},\ and\ \citenamefont {Uryū}}]{markakis2019timesymmetry}%
  \BibitemOpen
  \bibfield  {author} {\bibinfo {author} {\bibfnamefont {C.~M.}\ \bibnamefont
  {Markakis}}, \bibinfo {author} {\bibfnamefont {M.~F.}\ \bibnamefont
  {O'Boyle}}, \bibinfo {author} {\bibfnamefont {D.}~\bibnamefont {Glennon}},
  \bibinfo {author} {\bibfnamefont {K.}~\bibnamefont {Tran}}, \bibinfo {author}
  {\bibfnamefont {P.}~\bibnamefont {Brubeck}}, \bibinfo {author} {\bibfnamefont
  {R.}~\bibnamefont {Haas}}, \bibinfo {author} {\bibfnamefont {H.-Y.}\
  \bibnamefont {Schive}}, \ and\ \bibinfo {author} {\bibfnamefont
  {K.}~\bibnamefont {Uryū}},\ }\href@noop {} {\  (\bibinfo {year} {2019})},\
  \Eprint {http://arxiv.org/abs/1901.09967} {arXiv:1901.09967 [math.NA]}
  \BibitemShut {NoStop}%
\bibitem [{\citenamefont {O'Boyle}\ \emph {et~al.}(2022)\citenamefont
  {O'Boyle}, \citenamefont {Markakis}, \citenamefont {Da~Silva}, \citenamefont
  {Panosso~Macedo},\ and\ \citenamefont {Kroon}}]{OBoyle:2022yhp}%
  \BibitemOpen
  \bibfield  {author} {\bibinfo {author} {\bibfnamefont {M.~F.}\ \bibnamefont
  {O'Boyle}}, \bibinfo {author} {\bibfnamefont {C.}~\bibnamefont {Markakis}},
  \bibinfo {author} {\bibfnamefont {L.~J.~G.}\ \bibnamefont {Da~Silva}},
  \bibinfo {author} {\bibfnamefont {R.}~\bibnamefont {Panosso~Macedo}}, \ and\
  \bibinfo {author} {\bibfnamefont {J.~A.~V.}\ \bibnamefont {Kroon}},\
  }\href@noop {} {\  (\bibinfo {year} {2022})},\ \Eprint
  {http://arxiv.org/abs/2210.02550} {arXiv:2210.02550 [gr-qc]} \BibitemShut
  {NoStop}%
\bibitem [{\citenamefont {Markakis}\ \emph {et~al.}(2023)\citenamefont
  {Markakis}, \citenamefont {Bray},\ and\ \citenamefont
  {Zengino\u{g}lu}}]{Markakis:2023pfh}%
  \BibitemOpen
  \bibfield  {author} {\bibinfo {author} {\bibfnamefont {C.}~\bibnamefont
  {Markakis}}, \bibinfo {author} {\bibfnamefont {S.}~\bibnamefont {Bray}}, \
  and\ \bibinfo {author} {\bibfnamefont {A.}~\bibnamefont {Zengino\u{g}lu}},\
  }\href@noop {} {\  (\bibinfo {year} {2023})},\ \Eprint
  {http://arxiv.org/abs/2303.08153} {arXiv:2303.08153 [gr-qc]} \BibitemShut
  {NoStop}%
\bibitem [{\citenamefont {Da~Silva}\ \emph {et~al.}(2023)\citenamefont
  {Da~Silva}, \citenamefont {Panosso~Macedo}, \citenamefont {Thompson},
  \citenamefont {Kroon}, \citenamefont {Durkan},\ and\ \citenamefont
  {Long}}]{DaSilva:2023xif}%
  \BibitemOpen
  \bibfield  {author} {\bibinfo {author} {\bibfnamefont {L.~J.~G.}\
  \bibnamefont {Da~Silva}}, \bibinfo {author} {\bibfnamefont {R.}~\bibnamefont
  {Panosso~Macedo}}, \bibinfo {author} {\bibfnamefont {J.~E.}\ \bibnamefont
  {Thompson}}, \bibinfo {author} {\bibfnamefont {J.~A.~V.}\ \bibnamefont
  {Kroon}}, \bibinfo {author} {\bibfnamefont {L.}~\bibnamefont {Durkan}}, \
  and\ \bibinfo {author} {\bibfnamefont {O.}~\bibnamefont {Long}},\ }\href@noop
  {} {\  (\bibinfo {year} {2023})},\ \Eprint {http://arxiv.org/abs/2306.13153}
  {arXiv:2306.13153 [gr-qc]} \BibitemShut {NoStop}%
\bibitem [{\citenamefont {Da~Silva}(2024)}]{DaSilva:2024yea}%
  \BibitemOpen
  \bibfield  {author} {\bibinfo {author} {\bibfnamefont {L.~J.~G.}\
  \bibnamefont {Da~Silva}},\ }\href@noop {} {\  (\bibinfo {year} {2024})},\
  \Eprint {http://arxiv.org/abs/2401.08758} {arXiv:2401.08758 [gr-qc]}
  \BibitemShut {NoStop}%
\bibitem [{\citenamefont {Trefethen}(2000)}]{doi:10.1137/1.9780898719598}%
  \BibitemOpen
  \bibfield  {author} {\bibinfo {author} {\bibfnamefont {L.~N.}\ \bibnamefont
  {Trefethen}},\ }\href {\doibase 10.1137/1.9780898719598} {\emph {\bibinfo
  {title} {Spectral Methods in MATLAB}}}\ (\bibinfo  {publisher} {Society for
  Industrial and Applied Mathematics},\ \bibinfo {year} {2000})\BibitemShut
  {NoStop}%
\bibitem [{\citenamefont {Miguel}(2024)}]{Miguel:2023rzp}%
  \BibitemOpen
  \bibfield  {author} {\bibinfo {author} {\bibfnamefont {F.~S.}\ \bibnamefont
  {Miguel}},\ }\href {\doibase 10.1103/PhysRevD.109.104016} {\bibfield
  {journal} {\bibinfo  {journal} {Phys. Rev. D}\ }\textbf {\bibinfo {volume}
  {109}},\ \bibinfo {pages} {104016} (\bibinfo {year} {2024})},\ \Eprint
  {http://arxiv.org/abs/2308.03832} {arXiv:2308.03832 [gr-qc]} \BibitemShut
  {NoStop}%
\bibitem [{\citenamefont {Sheikh}(2022)}]{Sheikh:2022cud}%
  \BibitemOpen
  \bibfield  {author} {\bibinfo {author} {\bibfnamefont {L.~A.}\ \bibnamefont
  {Sheikh}},\ }\emph {\bibinfo {title} {{Scattering resonances and
  Pseudospectrum : stability and completeness aspects in optical and
  gravitational systems}}},\ \href@noop {} {Ph.D. thesis},\ \bibinfo  {school}
  {Institut de Math\'ematiques de Bourgogne [Dijon], France} (\bibinfo {year}
  {2022})\BibitemShut {NoStop}%
\bibitem [{\citenamefont {Boyd}(2001)}]{boyd2001chebyshev}%
  \BibitemOpen
  \bibfield  {author} {\bibinfo {author} {\bibfnamefont {J.~P.}\ \bibnamefont
  {Boyd}},\ }\href@noop {} {\emph {\bibinfo {title} {Chebyshev and Fourier
  spectral methods}}}\ (\bibinfo  {publisher} {Courier Corporation},\ \bibinfo
  {year} {2001})\BibitemShut {NoStop}%
\end{thebibliography}%
\bibliographystyle{apsrev4-1}

\end{document}